\documentclass[useAMS,usenatbib,fleqn]{mn2e}

\usepackage{bethmacros}
\usepackage{graphicx}
\usepackage{amssymb}
\usepackage{amsmath}

\usepackage{color}

\def\spose#1{\hbox to 0pt{#1\hss}}
\def\lta{\mathrel{\spose{\lower 3pt\hbox{$\mathchar"218$}}
     \raise 2.0pt\hbox{$\mathchar"13C$}}}
\def\gta{\mathrel{\spose{\lower 3pt\hbox{$\mathchar"218$}}
     \raise 2.0pt\hbox{$\mathchar"13E$}}}

\title[Local Photoionization Feedback - II]{Galaxy Formation with Local Photoionization Feedback - II. Effect of X-Ray Emission from Binaries and Hot Gas}

\author[Kannan et al.]{R. Kannan$^{1}$\thanks{Email: kannanr@mit.edu},\, M. Vogelsberger$^1$, G. S. Stinson$^{2}$, J. F. Hennawi$^2$, F. Marinacci$^1$, 
\newauthor{V. Springel$^{3,4}$, A. V. Macci\`o$^2$}
\vspace*{6pt}\\
$^1$Department of Physics, Kavli Institute for Astrophysics $\&$ Space Research, Massachusetts Institute of Technology, Cambridge 02139, MA, USA \\ 
$^2$Max-Planck-Institut f\"ur Astronomie, K\"onigstuhl 17, 69117, Heidelberg, Germany \\
$^3$Heidelberg Institute for Theoretical Studies, Schloss-Wolfsbrunnenweg 35, D-69118 Heidelberg, Germany \\
$^4$Zentrum f\"ur Astronomie der Universit\"at Heidelberg, ARI, M\"onchhofstr. 12-14, D-69120 Heidelberg, Germany}

\begin{document}
\maketitle
\label{firstpage}

\begin{abstract}

We study how X-rays from stellar binary systems and the hot intracluster medium
(ICM) affect the radiative cooling rates of gas in galaxies.  Our study uses a
novel implementation of gas cooling in the moving-mesh hydrodynamics code
\textsc{arepo}.  X-rays from stellar binaries do not affect cooling at all as
their emission spectrum is too hard to effectively couple with galactic gas.
In contrast, X-rays from the ICM couple well with gas in the temperature range
$10^4 - 10^6$ K. Idealised simulations show that the hot halo radiation field
has minimal impact on the dynamics of cooling flows in clusters because of the
high virial temperature ($\ga 10^7$K), making the interaction between the gas
and incident photons very ineffective. Satellite galaxies in cluster
environments, on the other hand, experience a high radiation flux due to the
emission from the host halo. Low mass satellites ($\la10^{12}\rm{M_\odot}$) in
particular have virial temperatures that are exactly in the regime where the
effect of the radiation field is maximal.  Idealised simulations of satellite
galaxies including only the effect of host halo radiation (no ram pressure
stripping or tidal effects) fields show a drastic reduction in the amount of cool gas
formed ($\sim 40\%$) on a short timescale of about $0.5$ Gyrs.  A galaxy merger simulation including all the other environmental quenching
mechanisms, shows about $20\%$ reduction in the stellar mass of the satellite
and about $\sim 30\%$ reduction in star formation rate after $1$ Gyr due to the
host hot halo radiation field.  These results indicate that the hot halo
radiation fields potentially play an important role in quenching galaxies in
cluster environments.
\end{abstract}


\section{Introduction}
 
The basic model of galaxy formation assumes that as dark matter hierarchically
collapses, gas flows into gravitational potential wells and forms stars at
their centers \citep{White1978}.  The gas converts gravitational potential
energy into thermal energy and consequently, the gas in the halo heats up to
the virial temperature, $T_{vir}$.  In the most massive galaxies, the
temperature becomes so large that the cooling time for the gas is longer than
the age of the Universe \citep{Rees1977}.  In this idealised picture, gas in
massive galaxies cannot cool to form stars.  However, galaxy haloes have high
density material in their centers, and since gas cooling is strongly dependent
on density, $\Lambda\propto n^2$, this dense central gas unstably overcools and
collapses to the halo center where it forms many more stars than are observed
\citep{Cole1991, Navarro1995, Benson2003}. 

To counteract this overcooling, galaxy formation models feed energy back from
stars \citep[e.g..][]{Navarro1996, Springel2003, Stinson2006, DV2008,
Vogelsberger2013, Fire2014} and active galactic nuclei (AGN)
\citep[e.g.,][]{Springel2005, DiMatteo2005N, Sijacki2007, Sijacki2009,
Booth2009}.  Stars deposit large amounts of energy through stellar winds,
supernovae explosions, and radiation \citep{Starburst99}.  Although feedback
from stars is able to regulate the star formation rate in low mass galaxies,
they are inefficient at high masses \citep{Kannan2014a}.  AGN feedback has been
invoked to solve the problem in high mass haloes ($\rm{M_{halo} \ga 
10^{12}M_\star}$).The details of how AGN couple to the gas in and around
galaxies is still uncertain, so modeling efforts have so far been necessarily
crude.

\citet{Springel2005}, for example,  used the energy from the high accretion
periods of AGN when they shine as quasi-stellar objects (QSOs, quasars) to heat
gas immediately surrounding the black hole and drive large scale winds.
However, this level of accretion requires that a  lot of cold gas is present at
the center of the galaxy,  which in turn leads to star formation. Even when
AGNs are accreting a small amount of gas, they drive spectacular radio jets
that are well collimated by magnetic fields and produce radio lobes tens of kpc
away from the black holes. Cosmological numerical simulations have cited these
jets as a form of feedback that is able to heat the entire gas halo to high
enough temperatures that prevents gas from cooling into the disk and forming
stars \citep{Sijacki2007}. However, these models do not address one of the
primary features of radio mode feedback, the strong directionality of the jet.
So neither of these models sufficiently explain how galaxies are quenched by
AGN feedback.

Thermal conduction is another mechanism which has been invoked to stabilise the
gas in clusters \citep{Zakamska2003, Voit2008, Voit2014N, Voit2011, Voit2015}.
The  large Chandra study of ICM entropy profiles (ACCEPT), showed that strong
AGN feedback is present only in clusters with low-entropy
cores~\citep{Cavagnolo2008, Cavagnolo2009}.  \citet{Voit2014N} have postulated
that the low entropy clusters have cooling times about the order of
precipitation times of the clusters, which is empirically seen to occur when
$\rm{t_{cool} \approx 10t_{ff}}$ \citep{Gaspari2011, Gaspari2012}. The clusters
with large entropy at their centers were shown to lie above the thermal
conduction balance solution, meaning that the cooling in the centers of these
clusters is balanced by thermal conduction heating by hot gas at larger radii.
One of the most important assumptions in this study is the value of the
conduction coefficient which is about $30 \%$ the canonical Spitzer
conductivity \citep{Spitzer1962}. Thermal conductivity can reach this high
value only if the magnetic field is chaotic over a wide range of length scales
(factor of 100 or more), as might happen with MHD turbulence
\citep{Narayan2001}. However, recent studies have shown that the reproduction
of observed sharp features in the ICM (cold filaments, bubbles) require that
the conduction coefficient be as low as $10^{-2} - 10^{-3}$ times the Spitzer
conductivity \citep{Gaspari2013, Arth2014}.   

\citet{Cantalupo2010} explored analytically the effect of X-ray radiation from
hot shock heated gas due to stellar winds and supernovae (SNe) explosions on
the cooling of halo gas.  The high-energy radiation from these sources was
shown to ionize metals and decrease the cooling rate of high-metallicity gas
considerably. \citet[][hereafter RK14b]{Kannan2014b} implemented this effect
using {\sc gasoline} \citep{Wadsley2004} and showed that in a cosmological
simulation of a Milky-Way galaxy the local radiation fields indeed reduce gas
cooling rate onto the disc of the galaxy thereby reducing the star formation.
This formalism will not quench galaxies because of the need to form stars in
order to reduce star formation. The radiation sources which can quench galaxies
need to be independent of the star formation in galaxies. Either the stellar
emission sources must have a time delay between the star formation event and
emission or they must be of a non-stellar origin.

\citet{Gnedin2012} created a general model for cooling in the presence of a
radiation field near a galaxy (including both stars and AGNs). They showed that
these radiation fields will significantly alter the cooling rates of gas in the
vicinity of a massive 'O' star and in galaxies containing highly luminous
quasars. Along these lines \citet{Vogelsberger2013} introduced a AGN feedback
mechanism, which modified the cooling of gas according to the incident
radiation field of the central AGN. In their simulations, the radiative
feedback is only effective during times when the AGN is strongly accreting and
shining brightly as a quasar, which again corresponds to times when stars are
forming. In the end, the only feedback that they found effective at maintaining
quiescent galaxies was the AGN radio-mode feedback that operates by heating the
halo gas to temperatures where its cooling time is too long to cool into the
disk to form stars.

Another avenue through which galaxies can quench is through environmental
effects such as ram pressure stripping \citep{Gunn1972, MCC2008, Font2008,
Kang2008, Weinmann2010} and tidal stripping \citep{Taylor2001, Donghia2010,
Chang2013}. These quenching mechanisms are invoked to explain the different
statistical properties of field and cluster galaxies, such as the prevalence of
ellipticals and lower star formation rates and hence redder colors. In many
prescriptions for ram pressure stripping the hot gas in the halo is assumed to
be stripped immediately after the satellite falls into the host, which leads to
a rapid decline in star formation rates \citep{Baldry2006, Weinmann2006,
Wang2007}.  More improved methods of stripping have been looked at, such as the
a more gradual stripping of the hot gas reservoir \citep{Font2008, Kang2008,
Weinmann2010, MCC2008}. However, even this formalism does not agree well with
the observational measurements and they generally overproduce the quenched
fraction of satellites as a function of distance from the cluster and
underproduce the fraction of elliptical galaxies in cluster environments
\citep{Guo2011, Weinmann2011}.

In this paper we explore the possibility that local X-ray radiation from X-ray
binaries (XRBs) and bremsstrahlung cooling radiation from hot gas present in
the haloes of galaxies can help to regulate star formation in galaxies. XRBs are
known to be the most prominent X-ray sources in a galaxy which is devoid of an
accreting black hole or large amounts of hot optically thin gas
\citep{Fabbiano2003}. Clusters of galaxies are the brightest extended X-ray
sources in the Universe, with their luminosities ranging from $10^{44-47}$
$\rm{erg \ s^{-1}}$ \citep{Vikhlinin2009, Pratt2009, Anderson2014}.  We build
on the RK14b model and account for these additional X-ray radiation fields
while calculating gas cooling rates in galaxy formation simulations. 

The paper is structured as follows. Section \ref{sec:ions} describes the
photoionization sources that we consider in this work. Section \ref{sec:calr}
outlines the approximations used in our radiative transfer approach, while
Section \ref{sec:ctc} describes the construction of the cooling table. Finally,
in Section \ref{sec:tpe} and \ref{sec:sim}, we present the results of our
implementation of the local photoionization feedback (LPF) model on a test gas
particle and on isolated galaxy simulations. Our conclusions are presented in
Section \ref{sec:conc}.

\begin{figure}
\begin{center}
\includegraphics[scale=0.47]{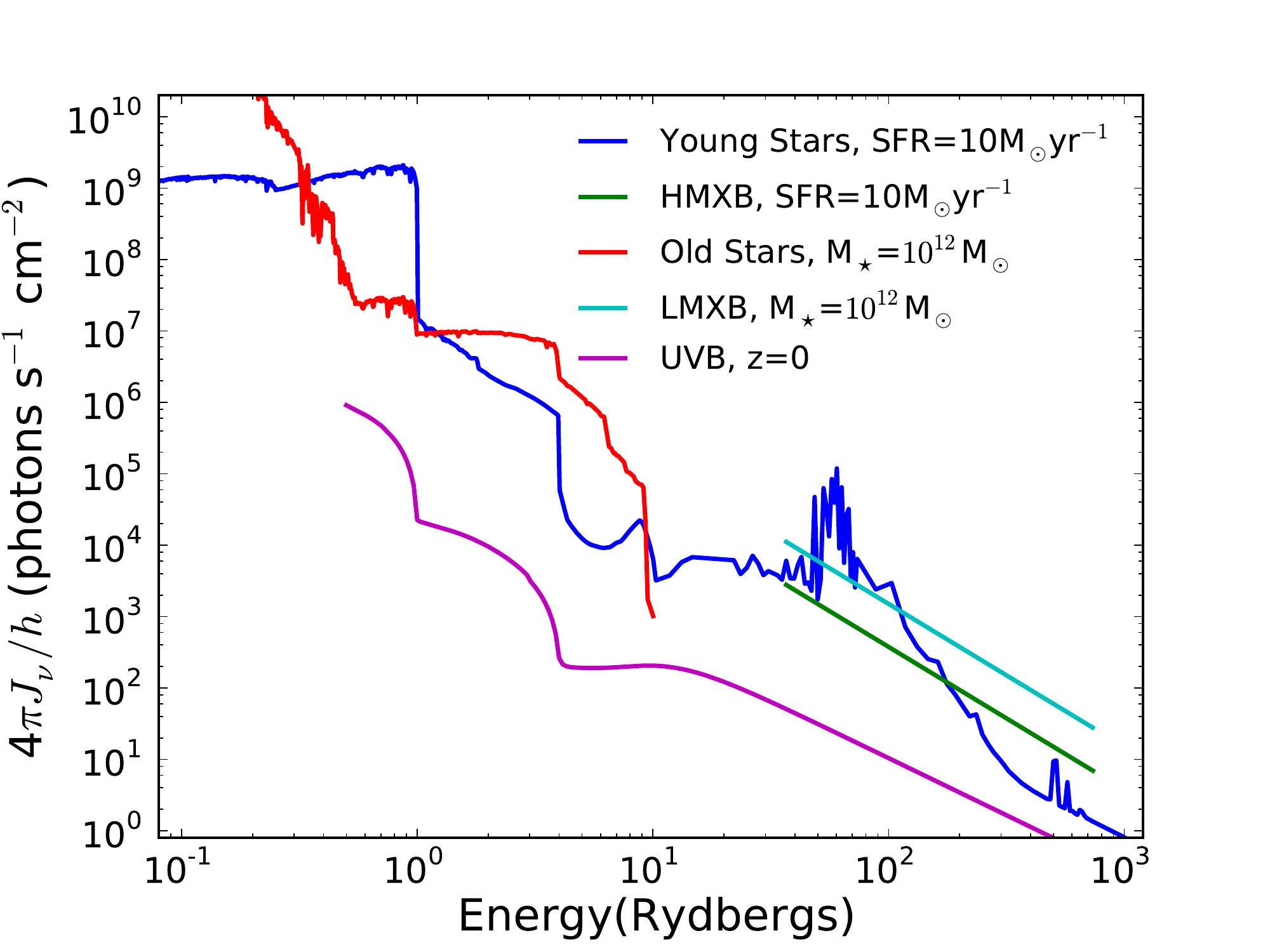}
\caption{The incident photon flux as a function of energy, from high (green
curve) and low (cyan curve) mass X-ray binaries compared to the radiation field
from star formation events or young stars (blue curve) and from old or post-AGB
stars (red curve).  These fluxes are reported  at $10$ \rm{kpc} from the
source. The radiation flux from both the HMXBs and the young stars are quoted
for a star formation rate of 10 M$_\odot$ yr$^{-1}$. The radiation flux from
both old stars and LMXBs are quoted for a stellar mass of $\rm{M_\star =
10^{12}M_\odot}$. For comparison, the estimate for the flux of the
extragalactic UV background at $z=0$ is also plotted (purple curve).}
\label{fig:spectra}
\end{center}
\end{figure}

\section{Ionizing sources}
\label{sec:ions}
There are several radiation sources that produce high energy ionizing photons.
RK14b modelled the effect of radiation from star formation events (or young
stars) and post-AGB (old stars) stars. In this paper we explore the emission
from X-ray binaries (XRBs) and the cooling radiation from the hot tenuous gas
around haloes. Throughout this work, the estimate for the extragalactic UV
background by \citet{CAFG2009}  is used. 

\subsection{X-Ray Binaries}
About a third of all stars form as part of a binary system \citep{Lada2006}.
The evolutionary timescales of stars comprising the binary system may be
different, leading to the collapse of one star earlier than the other. The
collapsed star then acts as an accretor of material from the donor main
sequence star, once the surface of the donor star crosses the inner Lagrange
point of the system \citep{Pac1971}. This infalling material dissipates
gravitational potential energy in the form of radiation. They are the most
prominent X-ray sources in galaxies which are devoid of an accreting black hole
or large amounts of hot optically thin gas \citep{Fabbiano2003}.  The XRB
population is divided into three categories depending on the mass of the donor
star: high mass XRBs ($\rm{M_\star} \ \ge 8 \ \rm{M_\odot}$), intermediate mass
XRBs ($1 \ \rm{M_\odot} < \rm{M_\star} < 8 \ \rm{M_\odot} $) and  low mass XRBs
($\rm{M_\star} \ \le 1 \ \rm{M_\odot}$). The X-ray emission is dominated by
high mass XRBs immediately after a star formation event, followed by a possible
contribution from the intermediate mass XRB population, and the late time
emission is dominated by low mass XRBs. The emission from and properties of,
intermediate mass XRBs is not well studied because there are very few XRBs in
this mass range in the Milky-Way galaxy \citep{Pod2002,Pfahl2003}. Therefore,
in this work we only consider the emission from the high and low mass XRB
populations only.

\begin{figure}
\begin{center}
\includegraphics[scale=0.47]{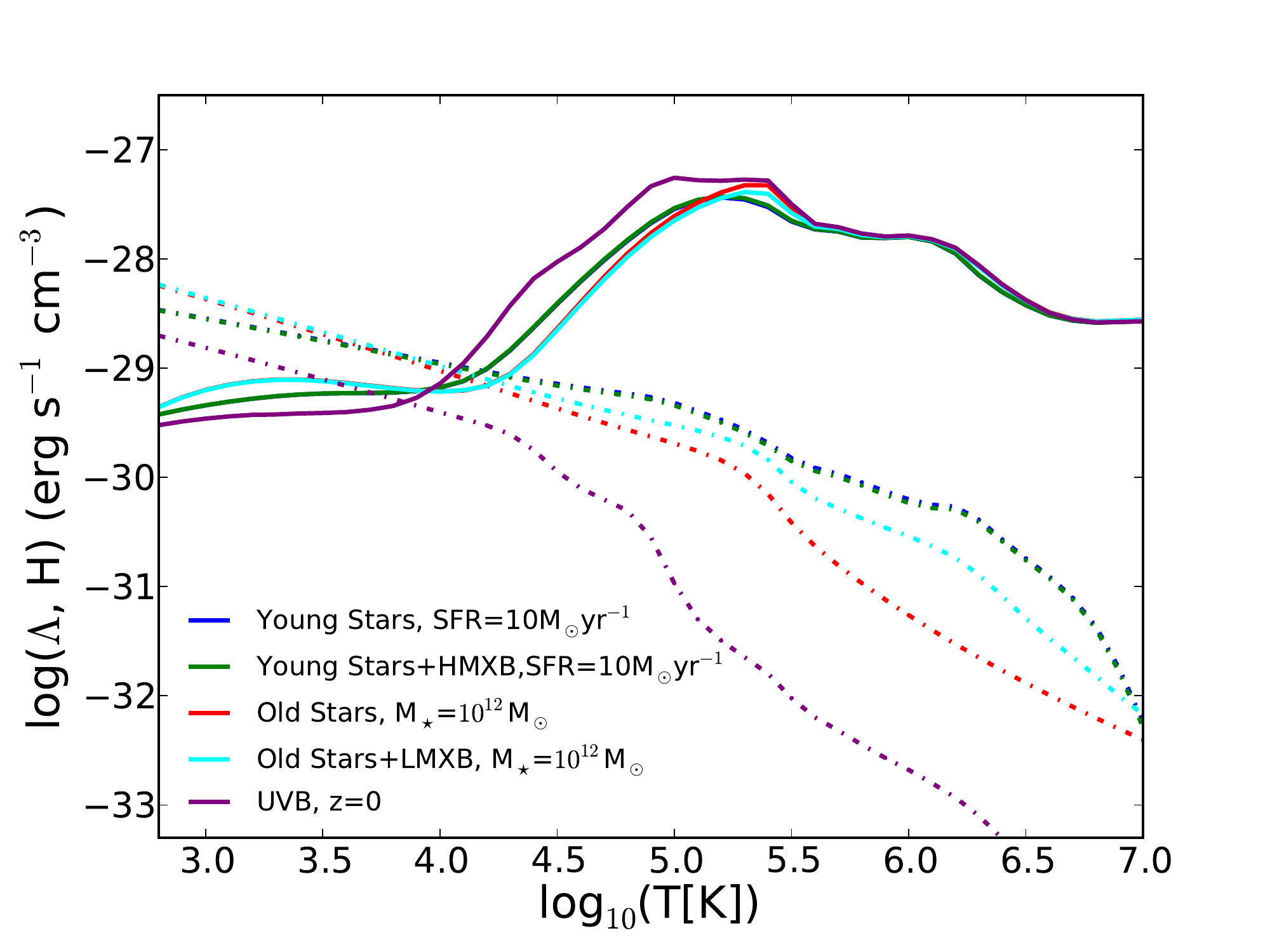}
\caption{The volumetric cooling (solid curves) and heating (dot dashed curves)
rates of gas under the influence of radiation from young stars and HMXBs (blue
curves), only young stars (green curves), old stars and LMXBs (cyan curves),
only old stars (red curves) and the extragalactic UV background (purple
curves). The gas is at a density of $n_H = 0.001$\rm{cm}$^{-3}$, and a
metallicity of $\rm{Z}_{\odot}$. The radiation flux for each component is equal
to the value quoted in Fig. \ref{fig:spectra}.}
\label{fig:cooling}
\end{center}
\end{figure}

\subsubsection{High Mass X-ray Binaries}
High mass XRBs (HMXBs) are binaries where the mass of the donor star is $\ga 8
\rm{M}_\odot$. Their lifetimes are determined by the nuclear timescale of the
high mass donor star, which is  $1 - 10 $ Myr. Thus, HMXB emission occurs a few
million years after star formation events and the total emission luminosity of
HMXBs follows the star formation rate of the galaxy.  Tight empirical relations
are observed between the X-ray luminosity of the HMXBs and the star formation
rate of the galaxy \citep{Grimm2003, Persic2004}. Recently \citet{Mineo2012}
presented the following relation for the luminosity of the HMXBs as a function
of the star formation rate 
\begin{equation}
L_{0.5-8\rm{keV}}^{\rm{\small HMXB}} = 2.61 \times 10^{39} \ \left(\frac{SFR}{1\rm{M}_\odot \rm{yr}^{-1}}\right)  {\rm erg} \ {\rm s}^{-1}.
\label{eq_hmxbsfr}
\end{equation}
Energies lower than $0.5$ keV are difficult to observe, because they are
absorbed in the interstellar medium of the host as well as our own Milky-Way
galaxy. We therefore model XRB emission in the hard X-ray regime only. In this
regime (above $0.5$ keV), the HMXB SED can be approximated by a power law: 
       \begin{equation}
  \frac{dI}{dE}  \propto E^{-2},
       \label{eq_xrbspecshape}
       \end{equation}
where $I$ is the intensity and $E$ is the energy of the emitted photon.  Using
these relations we can plot the SED of HMXBs as a function of the star
formation rate of the galaxy. Fig. \ref{fig:spectra} shows the HMXB emission
spectra corresponding to a SFR of $\rm{10 \ M_\odot \ yr^{-1}}$ (green curve).
For comparison, the SED from young stars (see RK14b for details) is shown for
the same SFR (blue curve). The fluxes quoted are at a distance of $10$ kpc from
the source. 

Fig.\ref{fig:cooling} shows the corresponding cooling (solid curves) and
heating (dashed curves) rates in presence of various radiation sources. The gas
parcel is at a density of $n_{\rm H} = 0.001 \rm{cm}^{-3}$ and $\rm{Z} =
\rm{Z}_\odot$ and at a distance 10 kpc from the source. The cooling curve is
calculated under the assumption that the gas in between the source and the gas
parcel is optically thin. Thus, the flux is inversely proportional to the
distance squared. The heating ($\rm{H}$) and cooling ($\Lambda$) curves of the
gas in the presence of the young stars (blue curves) radiation field only lies
exactly on top of the heating and cooling curves calculated in the presence of
young stars + HMXB radiation field (green curves).  In a star forming galaxy
the emission from young stars (i.e., blackbody emission from stars
and X-ray emission from shock heated gas) dominates the emission from HMXBs up
to $\sim 180$ Rydbergs (Fig. \ref{fig:spectra}).  Above this energy, the hard X-ray photons do not couple effectively with the gas because of low interaction
cross section.  We conclude that HMXBs, although an important source of X-rays
have a sub-dominant effect on the gas cooling rates in galaxies.

\subsubsection{Low Mass X-Ray Binaries}
Low Mass XRBs (LMXBs) are binaries where the mass of the donor star is $\la 1
\rm{M}_\odot$. The X-ray active phase of these systems are delayed by the
nuclear timescale of the low mass donor star or by the orbital decay timescale.
These timescales are $\sim 1-10$ Gyrs \citep{Verbunt1995}. The duration of the
X-ray active phase also has similar timescales \citep{Pod2002,Pfahl2003}. This
delay in X-ray active phase decouples the X-ray emission from the star
formation event. Instead, the luminosity of LMXB correlates well with the total
stellar mass of the galaxy \citep{Gilfanov2004}, and the relation is given by:

\begin{equation}
L_{0.5-8\rm{keV}}^{\rm{\small LMXB}}  = 1.00 \times 10^{40} \ \left(\frac{\rm{M}_\star^{tot}}{10^{11}\rm{M}_\odot}\right) {\rm erg} \ {\rm s}^{-1}.
\label{eq_lmxbsfr}
\end{equation}

The spectral shape of LMXB emission is assumed to follow the same functional
form as  Eq.~\ref{eq_xrbspecshape}. Fig.~\ref{fig:spectra} shows the LMXB
emission flux (cyan curve) for a galaxy with a total stellar mass of $10^{12}
\rm{M_\odot}$ at a distance of $10$ kpc from its center.   For comparison the
flux from post-AGB stars (red curve), under similar conditions is also shown.
The LMXBs contribution is primarily in the hard X-ray regime, while post-AGB
stars shine only in the UV. 

The corresponding cooling curves are plotted in Fig.\ref{fig:cooling}. The
cooling curves for a parcel of gas under the influence of radiation from just
post-AGB stars (red curve) and one with radiation from post-AGB and LMXBs
(cyan curve) are indistinguishable, except for a slight suppression of cooling
at $\rm{log}_{10}(\rm{T [K]})\sim 5.3$ in the presence of LMXB radiation field.
This implies that XRBs in general do not have a significant impact on the
cooling rates. This is due to a couple of factors, one, their emission fluxes
are sub dominant compared to the radiation sources already considered in RK14b,
i.e., young stars and post-AGB stars. And two, the hard X-ray photons
have very low interaction cross sections. If we could in fact include the XRB
emission at lower energies, then it could possibly have an impact on gas
cooling rates. However, modelling this part of the emission spectra has so far
proven to be quite challenging  \citep{Gilfanov2010N}.

\subsection{Hot Gas Emission}
\label{sec:hhe}
Early observations indicated that clusters are ubiquitous emitters in the X-ray
regime, with luminosities $\sim 10^{42-45} \rm{erg \ s^{-1}}$
\citep{Giacconi1972}. The X-ray emission is observed to be non time-varying and
extended, with the spatial extent matching the galaxy distribution
\citep{Kellogg1972, Elvis1976}.   These properties point to the fact that the
X-ray emission arises from  bremsstrahlung  or free-free cooling of the hot
(T$\approx 10^{7-8} \rm{K}$) tenuous (n$_{\rm H} \approx 10^{-3} \
\rm{cm^{-3}}$) intracluster medium (ICM). The thermal free-free volumetric
emissivity ($j_\nu^{ff}$) of a parcel of gas at temperature  $T$ is given by 
\begin{equation}
   j_{\nu}^{ff} = \Lambda_{\nu}(T) n_{e} n_{i}, 
   \end{equation}
   where
   \begin{equation}
   \Lambda_{\nu}(T) = 6.8 \times 10^{-38} T^{-1/2} e^{-h\nu/kT} Z^2 g_{ff},  
  \label{eq:bremms}
 \end{equation} 
where $g_{ff}$ is the free-free gaunt factor, $k$ is the Boltzmann constant,
$Z$ is the mean ionic charge, $h$ is the Planck constant and $n_e$ and $n_i$
are the electron and ion particle densities respectively.  The free-free
emission SED is independent of the frequency at low energies and has an
exponential cutoff at a characteristic frequency, determined by the temperature
of the gas. This cutoff frequency shifts towards higher values as the
temperature of the gas increases. The total flux on the other hand is a strong
function of the density of the emitting gas parcel $j_\nu^{ff} \propto \rho^2$.
Therefore, the temperature of the gas decides the hardness of the spectrum
while its density controls the overall luminosity.

\begin{figure}
\begin{center}
\includegraphics[scale=0.47]{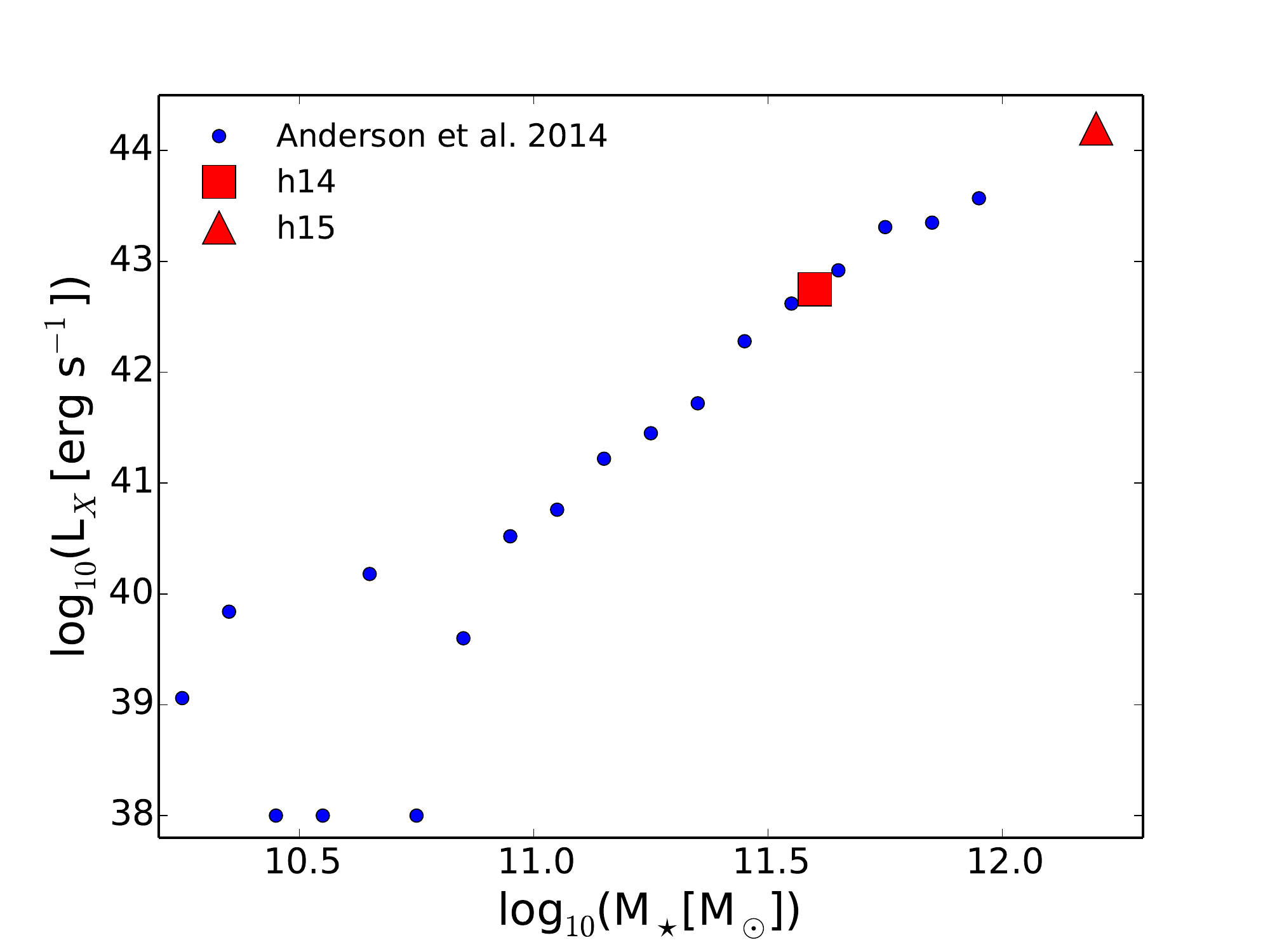}
\caption{X-ray (0.5-2.0 keV) luminosities of clusters.  The blue points are
observational measurements taken from \protect\cite{Anderson2014}. The red
square (h14) and triangle (h15) are the luminosities of the synthetic clusters
calculated using Eqs. \ref{eq:lnu},\ref{eq:lnutoto}. These luminosity
measurements are made between $0.15 -1.0 \ \rm{R}_{500}$.}
\label{fig:lum}
\end{center}
\end{figure}

In order to study the effect of the X-rays from clusters on gas cooling rates,
it is important to match their observed luminosities. The relation between the
X-ray luminosity and the mass of the host halo has been extensively studied all
the way from galaxy groups \citep{Bharadwaj2014} to cluster scales
\citep{Vikhlinin2009, Pratt2009, Planck2013, Wang2014}. More recently
\citet{Anderson2014} analyzed a sample of $\sim 250000$ locally brightest
central galaxies selected from the Sloan Digital Sky Survey (SDSS) and stacked
the X-ray emission from these haloes, as a function of the stellar mass of the
central galaxy, using data from the \textsc{ROSAT} All-Sky Survey.  They
constrain the $L_x - M_\star$ relation in a wide range of central galaxy
stellar masses shown in Fig. \ref{fig:lum} (blue points). These values are
quoted for hot gas emission between $0.15-1.0 \ \rm{R_{500}}$ of the cluster.

We construct two idealised haloes containing gas and DM, one of mass
($\rm{M_{halo}}$) $10^{14} \ \rm{M}_\odot$ (h14) and one of mass $10^{15} \
\rm{M}_\odot$ (h15). The DM density profile of the these haloes follows a
\citet{NFW1997} (NFW) distribution. The concentration parameter is set by the
$c_{vir}- M_{halo}$ relation given in \citet{Dutton2014}. The gas fraction of
the haloes is chosen to match observations \citep{Giodini2009, Gonzalez2013,
Sanderson2013}. The gas in the halo is placed in hydrostatic equilibrium with
the DM and the temperature of the gas is set to the virial temperature of the
halo.  The properties of these haloes are shown in Table~\ref{tab:clprop}.
These haloes are then evolved, using {\sc arepo}, with cooling, gravity and
hydrodynamics turned on for $200$ Myrs. The luminosities are then
calculated using the following relations (see also Eq.~$2$ of
\citealt{Crain2010}):
  \begin{equation}
  \begin{split}
   L_{\nu,a} &=j_{\nu,a} V_a \\ 
             &= \Lambda_{\nu,a}(T_a) n_{e,a} n_{i,a} V_a \\
             &= \Lambda_{\nu,a}(T_a) \frac{X_e}{(X_e+X_i)^2}\left( \frac{\rho_a}{\mu m_p} \right)^2 V_a \\
             &= \Lambda_{\nu,a}(T_a) \frac{X_e}{(X_e+X_i)^2} \frac{\rho_a}{\mu m_p} \frac{m_a}{\mu m_p}, \\
               \label{eq:lnu}
   \end{split}
 \end{equation}
where $V_a$ is the volume of the ${\rm a}^{th}$ gas cell, $\rm{X_e = n_e/n_H}$,
$\rm{X_i = n_i/n_H}$, $\rho_a$ is the density of the cell, $m_p$ is the proton
mass and $m_a$ is the mass the gas cell. The electron and ion densities of any
given gas cell are calculated assuming a primordial composition.  The total
X-ray luminosity is then given by 
\begin{equation}
L = \sum_a \int_{0.5 \ \rm{keV}}^{2 \ \rm{keV}} L_{\nu, a} d\nu.
\label{eq:lnutoto}
\end{equation} 
The  luminosity calculated using these relations for the clusters h14 (red
square) and h15 (red triangle) is shown in comparison to the observational data
points from \citet{Anderson2014} in Fig.~\ref{fig:lum}. The stellar mass of the
haloes in question are determined by the $\rm{M}_\star - \rm{M_{halo}}$ relation
given by \citet{Kravtsov2014}.  The X-ray luminosity of the synthetically
constructed clusters matches quite well the observationally derived values. 

\begin{figure}
\begin{center}
\includegraphics[scale=0.47]{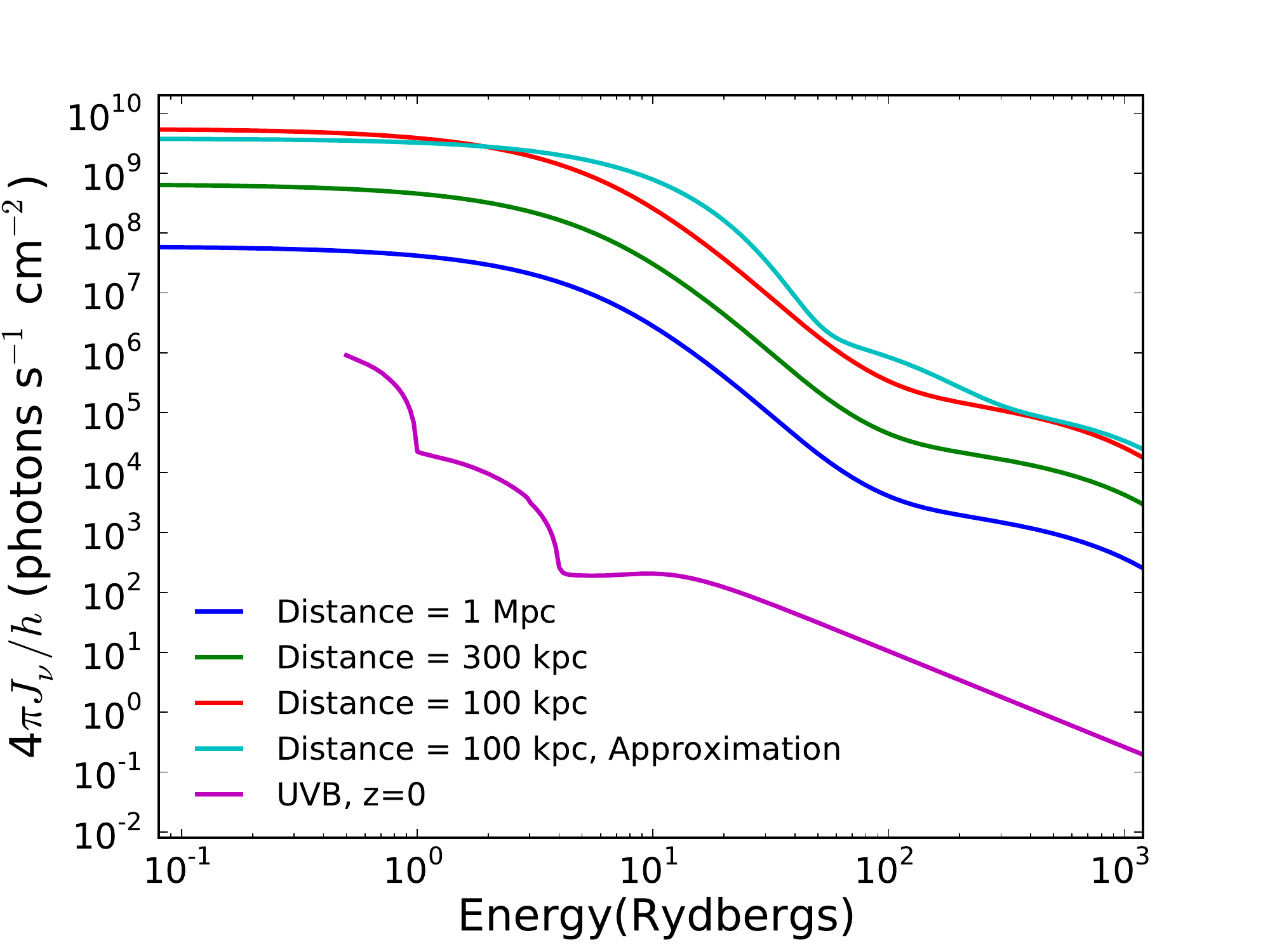}
\caption{The incident photon flux as a function of energy due to bremsstrahlung
emission from the hot ICM, at various distances from the center of the h15
cluster. The cyan curve is the flux at $100$ kpc using the three temperature
approximation described in Section \ref{sec:calr}. The other curves show the
fluxes at $100$ kpc (red curve), $300$ kpc (green curve) and at $1$ Mpc
distance from the center of the cluster, calculated using the exact relations
given by Eq. \ref{eq:lnu}. For comparison, the estimate for the flux of the
extragalactic UV  background at $z=0$ is also plotted (purple curve).}
\label{fig:spectra_hh}
\end{center}
\end{figure}

Using the relations given by Eq. \ref{eq:lnu}, we then calculate the hot gas
free-free emission fluxes at $100$ kpc (red curve), $300$ kpc (green curve) and
$1$ Mpc (blue curve) from the center of the h15 cluster (shown in
Fig.~\ref{fig:spectra_hh}). The SEDs show the characteristic shape of free-free
emission, with more than one temperature component contributing to the overall
shape. At $z=0$, the hot halo cooling flux is much larger than the contribution
from the extragalactic UV background (purple curve).

\begin{figure}
\begin{center}
\includegraphics[scale=0.47]{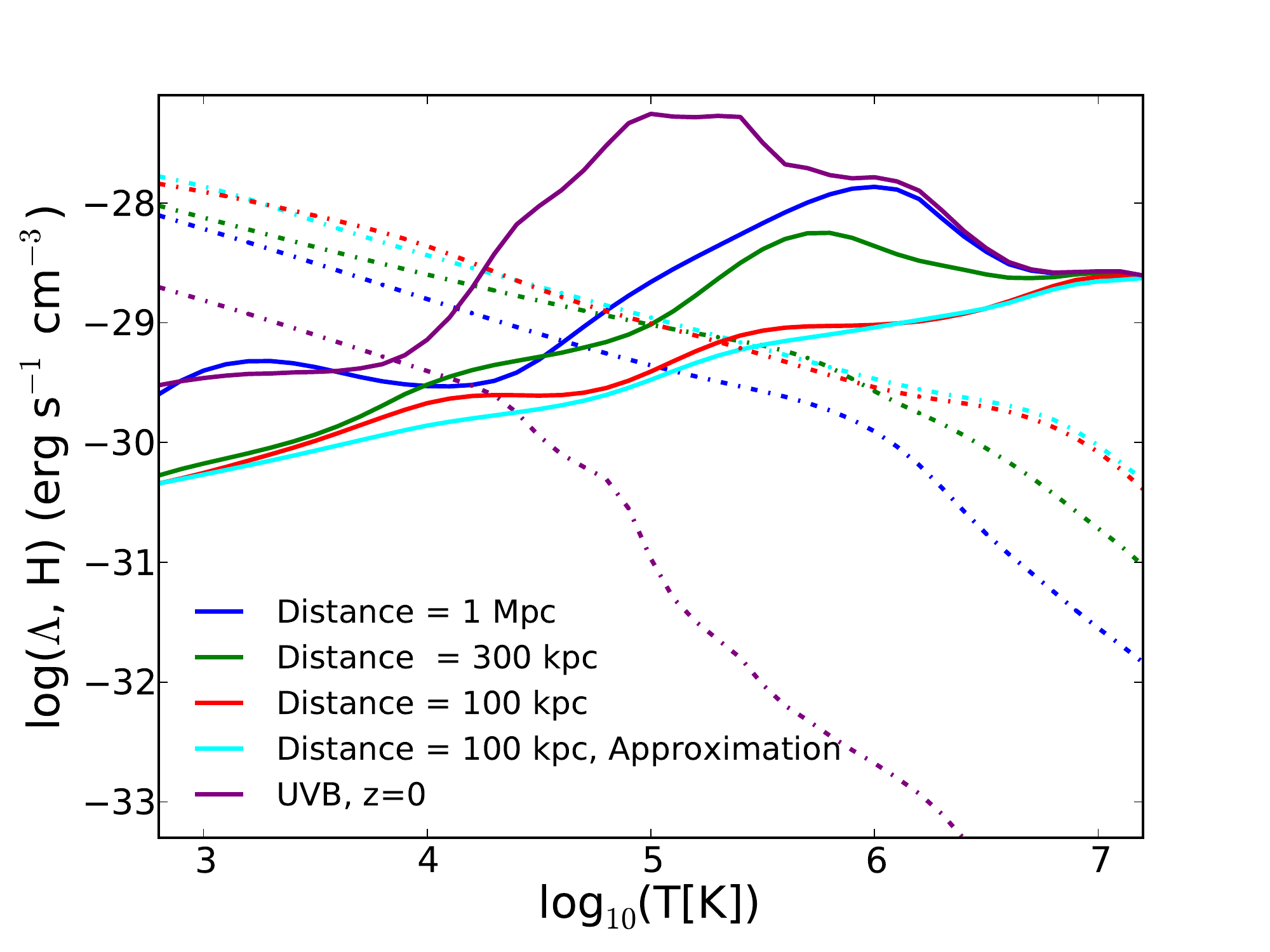}
\caption{The volumetric cooling (solid curves) and heating (dot dashed curves)
rates for a gas parcel under the influence of hot halo radiation field from the
h15 cluster at a distance of $100$ kpc (red curves), $300$ kpc (green curves)
and $1$ Mpc (blue curves). We also show the cooling and heating rates for an
SED at $100$ kpc calculated using the three temperature approximation (cyan
curves) described in Section \ref{sec:calr}. For comparison the cooling and
heating rates in the presence of an extragalactic UV background (purple curves)
at $z=0$ is also shown. The gas is at a density of $n_H = 0.001$\rm{cm}$^{-3}$,
and a metallicity of $\rm{Z}_{\odot}$. }
\label{fig:cooling_hh}
\end{center}
\end{figure}

The cooling curves of a gas parcel of $\rm{n_H = 0.001 cm^{-3}}$ and $\rm{Z =
Z_\odot}$ under the influence of the h15 cluster hot-halo radiation fields are
shown in Fig.~\ref{fig:cooling_hh}. The reduction of cooling between $10^4$ K
and $10^6$ K is quite drastic even at a distance of $1$ Mpc, with Fe  being the
only ionic species not completely photoionized. By comparison, a gas parcel at
$100$ kpc is completely ionized. The only cooling contribution is from
bremsstrahlung emission. The equilibrium temperature, i.e., the temperature at
which the heating and cooling rates of the gas parcel are equal, under these
conditions also increases to about $ 3 \times 10^5$ K.

Another important point to note is that the cooling rate of gas with
temperatures greater than $\sim 10^{6.5}$ K does not change and seems independent
of the incident radiation field. The high temperature plasma is completely
thermally ionized and the only cooling mechanisms are bremsstrahlung and
Compton cooling. As the virial temperature of  clusters are also above $\sim
10^{7}$K, most of the ICM will be optically thin to the X-ray emission, which
is in agreement with observations \citep[and references within]{Sarazin1986}.
Therefore, the radiation field will not affect a large fraction of the cluster
gas, but it will certainly reduce the cooling rate of any gas in the cluster
with relatively low temperatures, either present in the form of cooling flows
or as part of satellite galaxies. We can therefore conclude that the hot halo
radiation fields will have a non-negligible effect on gas cooling rates in
clusters.

\begin{table}
 \centering
  \begin{tabular}{lllllccl}
  \hline
  Halo                        \!\!\!\!\!\!\!\!     & 
  $\rm{M}_{halo}$[M$_\odot$]  \!\!\!\!\!\!\!\! & 
  $N_{\rm DM}$                \!\!\!\!\!\!\!\!                     & 
  $N_{\rm gas}$               \!\!\!\!\!\!\!\!   & 
  $c_{\rm vir}$               \!\!\!\!\!\!\!\!                     & 
  $\epsilon_{DM}$[kpc]        \!\!\!\!\!\!\!\!                     & 
  $\epsilon_{gas}$[kpc]       \!\!\!\!\!\!\!\!                     & 
  $f_{\rm gas}$                                   \\
 \hline
 h15                          \!\!\!\!\!\!\!\!                              & 
 $10^{15}$                    \!\!\!\!\!\!\!\!                    & 
 $50000$                      \!\!\!\!\!\!\!\!                      & 
 $100000$                     \!\!\!\!\!\!\!\!                     & 
 $5.21$                       \!\!\!\!\!\!\!\!                     & 
 $7.57$                       \!\!\!\!\!\!\!\!                     & 
 $2.92$                       \!\!\!\!\!\!\!\!                     & 
 $0.10$                                            \\
 h14                          \!\!\!\!\!\!\!\!                              & 
 $10^{14}$                    \!\!\!\!\!\!\!\!                    & 
 $50000$                      \!\!\!\!\!\!\!\!                      & 
 $100000$                     \!\!\!\!\!\!\!\!                     & 
 $6.56$                       \!\!\!\!\!\!\!\!\!\!                    & 
 $3.57$                       \!\!\!\!\!\!\!\!                     & 
 $1.27$                                            & 
 $0.08$                                            \\
 h511                         \!\!\!\!\!\!\!\!                             & 
 $5 \times 10^{11}$           \!\!\!\!\!\!\!\!           & 
 $20000$                      \!\!\!\!\!\!\!\!                      & 
 $40000$                      \!\!\!\!\!\!\!\!                     & 
 $8.0$                        \!\!\!\!\!\!\!\!                     & 
 $0.82$                       \!\!\!\!\!\!\!\!                     & 
 $0.37$                       \!\!\!\!\!\!\!\!                     & 
 $0.15$                                            \\
  h15hr                        \!\!\!\!\!\!\!\!                             & 
 $10^{15}$           \!\!\!\!\!\!\!\!           & 
 $2.5 \times 10^6$                      \!\!\!\!\!\!\!\!                      & 
 $5 \times 10^6$                      \!\!\!\!\!\!\!\!                     & 
 $5.21$                        \!\!\!\!\!\!\!\!                     & 
 $1.30$                       \!\!\!\!\!\!\!\!                     & 
 $0.60$                       \!\!\!\!\!\!\!\!                     & 
 $0.10$                                            \\
  h212lr                         \!\!\!\!\!\!\!\!                             & 
 $2 \times 10^{12}$           \!\!\!\!\!\!\!\!           & 
 $5000$                      \!\!\!\!\!\!\!\!                      & 
 $10000$                      \!\!\!\!\!\!\!\!                     & 
 $10.0$                        \!\!\!\!\!\!\!\!                     & 
 $1.30$                       \!\!\!\!\!\!\!\!                     & 
 $0.60$                       \!\!\!\!\!\!\!\!                     & 
 $0.15$                                            \\
\hline
\end{tabular}
\caption{Table listing the parameters of the simulated galaxies}
\label{tab:clprop}
\end{table}

\section{Calculating the Radiation Field}
\label{sec:calr}
In the previous section the different sources of X-ray radiation were
enumerated and the hot ICM cooling radiation was shown to be important. In
RK14b we used an optically thin approximation, with a fixed escape fraction for
low energy photons. This allowed us to mimic absorption and at the same time
make the radiative transfer problem computationally feasible. 

The low energy part of the XRB spectrum is not observable due to absorption by
the disk of the host as well as our own Milky-Way (MW) galaxy.  This
necessitates that we only model XRB emission in the hard X-ray regime. The
interaction cross section for the hard X-ray photons is small and hence their
escape fraction from the disk of the galaxy is expected to be unity. Therefore
the optically thin approximation holds for the XRB radiation fields.
Additionally, we have argued in Section \ref{sec:hhe} that most of the hot
inter cluster gas is optically thin to the radiation field. This approximation
becomes invalid at the centers of the clusters, where the gas becomes dense
enough to self-shield against the incident radiation field. This effect is
modelled by assuming that gas above a certain density no longer sees the
incident radiation field (see Section \ref{sec:sim} for more details).
Therefore the optically thin approximation is valid for both radiation fields
considered in this paper making radiation transfer a simple inverse distance
squared problem.

\subsection{Three temperature approximation}

The RK14b model, made a couple of simplifying assumptions in order to quantify
the effect of local radiation fields on gas cooling rates in galaxy formation
simulations. One, except for absorption at the source, the gas is
assumed to be optically thin at all wavelengths and  two, the shape of each
contributing component (young stars and post-AGB stars) to the overall SED  does
not vary with the underlying properties of the component.  For example, the
model assumes that the SED from old stars is constant in shape irrespective of
the age of the stellar population (as long as $\rm{t_\star \ge 200 \ Myrs}$).
This made sure that the spectrum of each contributing component does not change
shape, rather only the overall normalisation of the spectrum varied according
to the availability of and distance from the source. Therefore the
photoionization ($\Gamma$) and photo heating ($\epsilon$) rates of each atomic
species `i' present in the gas, were just the sum of the individual components
`c', weighted according to their relative flux contributions `$\phi$' 
\begin{equation}
(\Gamma_{i}, \epsilon_{i}) = \sum_c \phi_c (\Gamma_{c,i}, \epsilon_{c,i}).
\label{eq:comp}
\end{equation}
This same method can be used for the XRB emission, because they scale exactly
in the same way as the young stars SED (which scales with SFR) and and emission
from post-AGB stars (which scales with the total stellar mass of the galaxy)
considered in RK14b. 

However, it is more difficult to model the emission from the hot gas. As
discussed in Section \ref{sec:hhe}, the temperature of the emitting gas parcel
controls the hardness of the spectra while the density controls the overall
flux. The emitted spectra of a cluster will vary depending on the gas
temperature and density distribution within the cluster. Therefore, in addition
to the overall normalisation, the shape of the SED also changes. The hot halo
emission, therefore, cannot be quantified as one of the components of Eq.
\ref{eq:comp}, instead one would theoretically need an infinite number of
components, spanning the entire temperature range of the ICM.  Every component
of the hot halo radiation field will contribute an additional dimension to the
cooling table. 

Observationally, the most dominant sources of hot gas emission are galaxy
groups and clusters. In lower mass haloes the hot halo emission is orders of
magnitude lower and sub-dominant with respect to other X-ray sources like XRBs
\citep{Bogdan2013b, Bogdan2013a, Anderson2014, Bogdan2015}. Since galaxy
groups/clusters usually have virial temperatures $T>10^{6.5}$K, we consider
only cooling radiation from relatively hot gas; i.e. gas with
$\rm{log_{10}(T[K])} > 5.5$. We further sub divide this hot gas into three
temperature regimes, $10^6$K, $10^7$K and $10^8$K, with a spread of $0.5$ dex
in each direction around each temperature bin. The gas particles in each bin is
then assumed to have the representative temperature of that bin.  The flux from
hot halo emission at any point `p' is then approximated as 
\begin{equation}
F_{\nu, p}^{ff} = \sum_{t} \Lambda_{\nu}(10^{t}) \frac{X_e}{(X_e+X_i)^2} \frac{1}{4\pi}\phi_{Tt,p}  \ , \   t \in [6,7,8], \\
\label{eq:3tempapprox}
\end{equation}
where
\begin{equation}
\phi_{Tt,p}  = \sum_a \frac{\rho_{a} m_{a}}{|\vec{r}_{a} - \vec{r}_p|^2 \mu^2 m_p^2 } \\
\forall     \ \rm{particles}   \  \rm{`a'},\\
\end{equation}
such that 
\begin{equation}
t-0.5 < \rm{log_{10}(T_{a}[K])} \le t+0.5.
\end{equation}

This approximation reduces the number of components of the hot halo emission
radiation to three. The SED of each component has a fixed shape due to the
fixed temperature and the normalisation just depends on the density times the
mass of the particles within that temperature range. To check the validity of
this approximation we plot, in Fig. \ref{fig:spectra_hh}, the radiation flux at
a point $100$ kpc from the center of the h15 cluster using the approximation
(cyan curve) compared to the  flux at the same point using exact relations (red
curve) described in Section \ref{sec:hhe}. The approximated SED matches quite
well with the actual SED, with slight differences especially in the regime
where the SED changes shape, around $100$ Rydbergs. The corresponding cooling
curve is plotted in Fig. \ref{fig:cooling_hh}. The heating rates (dot dashed
lines) lie on top of each other, while there seems to be a small difference to
the cooling component. This difference is however of the order of $0.05-0.1$
dex. Compared to the computational simplicity it offers, the errors induced by
the approximation are quite small.

\subsection{Propagating the radiation field}

Now that we have found an effective way to quantify XRBs and hot halo emission,
we need to come up with a computationally efficient method to propagate the
radiation field over the whole simulation volume. We have argued that the
radiation fields considered in this work, for all practical purposes, can be
assumed to be optically thin. This allows us to quantify the flux from each
component in the parameter `$\phi$', which is proportional to the availability
of the sources and inversely proportional to square of the distance between the
source and the sink. 

We use the same method described in RK14b to calculate the flux of the
radiation field; i.e. we tag on to the gravity tree of the simulation code.
The tree structure partitions the mass distribution into a hierarchy of
localised regions. The force from nearby particles is calculated accurately,
while the distant regions are explored more coarsely by treating them as single
massive pseudo-particles, centered on the center of mass of the corresponding
region. In RK14b we implemented the LPF model in {\sc gasoline}
\citep{Wadsley2004} which used a k-D (binary) tree structure \citep{Stadel2001}
in which bisections are done recursively through their longest axis of the
cells, whereas, in this paper we implement the model in {\sc arepo}
\citep{Springel2010} which uses the same oct-tree \citep{BarnesHut1986}
structure as {\sc Gadget-2} \citep{Springel2005}. Although these codes differ
in the actual construction of the gravity tree, they are the same for our
purposes.

One very simple improvement we have made, while tagging on to the gravity tree
in {\sc Arepo} is to consider an effective center mass of the node while
calculating the distance between the node and the sink position.  The effective
center of mass of a radiation field component `c' in any node is defined as the
center of effective mass of all the particles in that node that contribute to
that particular component.  The effective mass of a component in turn is
defined as quantity that controls the normalisation of the SED of that
component.  For example, effective mass of the new stars SED component is just
the mass of stars less than $10$ Myr old (see RK14b for more details), while
the effective mass of the hot halo emission components is the the mass times
the density of the gas particles. This assures that the effective center of
mass of a component in the node is close to the point in the node where the
contribution to that particular component is maximum.

\section{Cooling table creation}
\label{sec:ctc}
We tabulate the gas cooling rates under various gas densities, temperatures and
incident radiation fields. The total gas cooling rate is divided into
primordial, metal and Compton cooling (Eq. 15 of RK14b). The primordial and
Compton cooling rates are calculated on-the-fly (see \citealt{Vogelsberger2013}
for more details). The metals are assumed to be in ionization equilibrium and
their heating and cooling rates are tabulated across a range of physical
conditions using CLOUDY. This look-up table is then used in simulations.

Above $z=9$, the gas is assumed to be of purely primordial composition,
negating the need for a metal cooling table. From $z=9$ to $z=3$ the cooling
table has five dimensions corresponding to the temperature, density, flux from
new stars ($\rm{t_\star \le 10 \ Myrs}$), flux from old stars ($\rm{t_\star \ge
200 \ Myrs}$) and redshift components. At $z<3$ the table has eight dimensions,
the additional dimensions corresponding to the three hot halo emission
components, which start to become dominant only at low redshifts once
sufficiently massive haloes start forming. 

The density ranges from $10^{-9}$ to $10^4$ ${\rm cm}^{-3}$ with a spacing of
1.0 dex in log space. The temperature ranges from $10^2$ to $10^9$ K with a
resolution of 0.1 dex. $\phi_{SFR}$, which quantifies radiation from young
stars and the HMXBs, varies from $10^{-5}$ to $10^3$ $\rm{M_\odot \ yr^{-1} \
kpc^{-2}}$. $\phi_{os}$, which quantifies the radiation field from post-AGB
stars and LMXBs, varies from $10^6$ to $10^{12}$ $\rm{M_\odot \ kpc^{-2}}$. The
spacing for both these $\phi$'s is 1.0 dex.  All three hot halo components,
$\rm{\phi_{T6}, \ \phi_{T7} \ \&  \ \phi_{T8}}$, ranges from $10^{17.5}$ to
$10^{23.5}$ $\rm{cm^{-5}}$, with a spacing of 2.0 dex. The redshift dimension
that accounts for the change in  UV background ranges from $0.0$ to $9.0$ with
a resolution of 0.5.

\begin{figure*}
\begin{center}
\includegraphics[scale=0.65]{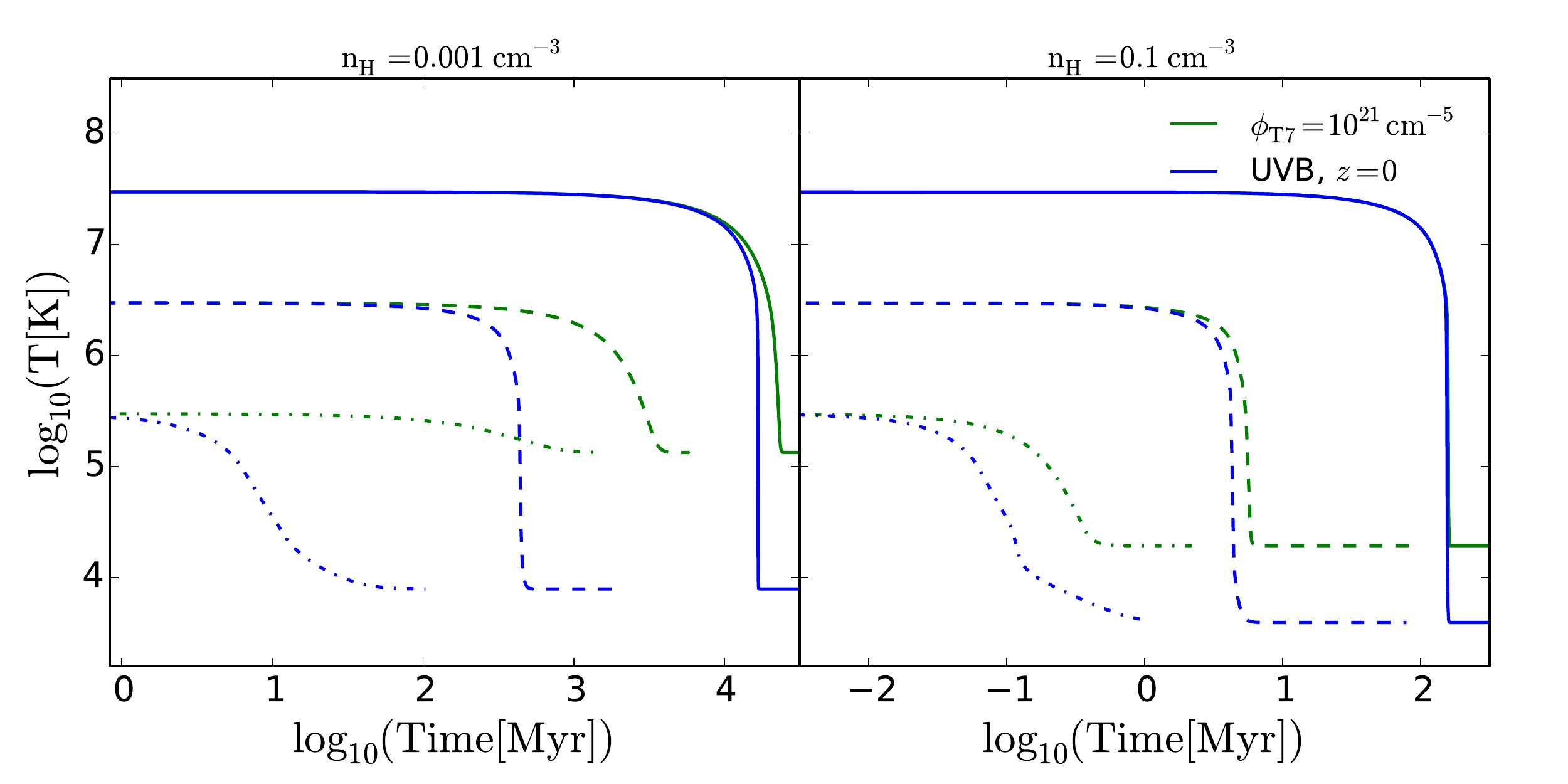}
\caption{The temperature evolution of gas parcels with densities $\rm{n_H =
0.001 \ cm^{-3}}$ (left panel) and $\rm{n_H = 0.1 \ cm^{-3}}$ (right panel) and
temperatures $3 \times 10^7$K (solid curves), $3 \times 10^6$K (dashed curves)
and $3 \times 10^5$K (dot dashed curves). Each gas parcel is subjected to two
different radiation fields, one which contains hot halo radiation field of
strength $\phi_{\rm{T7}} = 10^{21} \ \rm{cm^{-5}}$ plus the extragalactic UV
background at $z=0$ (green curves) and other which contains only the
extragalactic UV background flux at $z=0$ (blue curves).}
\label{fig:tpe}
\end{center}
\end{figure*}

To create the table, cooling and heating rates are calculated at every point of
phase space for solar and primordial metallicity gas using CLOUDY.  The
difference between the solar and primordial metallicity values are stored as
the heating and cooling rates due to metals only.

\section{Test particle evolution}
\label{sec:tpe}

As a first test of the cooling implementation in  {\sc arepo}, we calculate the
temperature evolution of isolated gas cells with and without hot halo radiation
fields (Fig. \ref{fig:tpe}). We consider gas cells of density of $\rm{n_H =
0.001 \ cm^{-3}}$ (left panel)  and  $\rm{n_H = 0.1 \ cm^{-3}}$ (right panel),
metallicity   $\rm{Z = Z_\odot}$ and starting temperatures  $3 \times 10^7$K
(solid curves), $3 \times 10^6$K (dot dashed curves) and $3 \times 10^5$K (dot
dashed curves). The temperature evolution of these six gas cells under the the
influence of a hot halo radiation field of strength $\phi_{\rm{T7}} = 10^{21} \
\rm{cm^{-5}}$  plus the UV background  at $z=0$ (green curves) and under the
influence of a UV background radiation field at $z=0$ only (blue lines) are
shown. 

The effect of the radiation field on the cooling rates of the gas cells is
highly dependent on both the starting temperature and density of the cell. At a
particular density, the initial temperature of the gas decides the
effectiveness of the radiation fields. The temperature evolution of the $3
\times 10^7$ K gas cell (at both densities) is barely effected.   The time it
takes for this high temperature gas cell to cool from $3\times 10^7$K to
$10^6$K is much longer ($\sim 20$ Gyrs for $\rm{n_H = 0.001 \ cm^{-3}}$) than
to cool from  $10^6$K to the equilibrium temperature ($\sim 1.7$ Gyrs for
$\rm{n_H = 0.001 \ cm^{-3}}$). Therefore, even if the radiation fields
drastically increases the cooling time from $10^6$K to $10^4$K (as seen in
Section \ref{sec:hhe}), the total cooling time will not change by a lot. 

Another notable change between the temperature evolution of gas cells with and
without the hot halo radiation fields is the increase in the minimum
temperature for each particle from $\sim10^4$K to $\sim10^5$K.  The minimum
temperature reflects the equilibrium temperature the radiation field sets.
While $10^5$ K is a higher temperature than that at which stars form, it is not
high enough to support against collapse with thermal pressure in cluster environments.  As we
will see in next section, $10^5$K gas collapses into a center that eventually
reaches a high enough density to cool regardless of the radiation field.

The cooling rates of the low temperature gas parcels ($3 \times 10^6$K \ \& \
$3 \times 10^5$K) on the other hand are reduced quite drastically. In the
presence of the hot halo radiation field the equilibration timescale is about
$7$ times higher for $3 \times 10^6$K gas parcel and $25$ times higher for $3
\times 10^5$K gas parcel at $\rm{n_H = 0.001  \ cm^{-3}}$.  In galaxies with
$T_{vir} < 10^{6.5}$K, the cooling and star formation timescales are inherently
smaller. Therefore an increase in the cooling time of the gas particle will
definitely affect the star formation rates of the galaxy.  This implies that
any galaxy that is close to or is falling into the cluster will experience
substantial quenching, due to the host halo radiation field. 

The density of the gas also plays a major role in its cooling rates. As we
discussed in Section \ref{sec:hhe}, the volumetric cooling rate is proportional
to the square of the density, which in turn means that the cooling time of the
gas parcel will be inversely proportional to the density of the gas
($\rm{t_{cool} = 3nk_BT/2\Lambda}$). This affect can clearly be seen in Fig.
\ref{fig:tpe}, where the cooling times of gas cells with $\rm{n_H = 0.1 \
cm^{-3}}$, irrespective of their starting temperature, is about a $100$ times
shorter than the cells with $\rm{n_H = 0.001 \ cm^{-3}}$. This means that
galaxies with higher mean gas densities will have a larger cooling flow. 

Additionally the effect of the radiation fields will also reduce at higher
densities. We can define an ionization parameter (U) for the gas parcel as 
\begin{equation}
U = \frac{4\pi}{h c n}   \int_{\nu_T}^\infty \frac{J_\nu}{\nu} d\nu,
\label{eq:ip}
\end{equation}
where $c$ is the speed of light, $h$ is the Planck constant, $n$ is the
particle density and $\nu_T$ is the threshold frequency for ionization (13.6eV
for hydrogen atom).  This parameter is basically the ratio between the number
density of ionizing photons incident on the gas cell to the number density
ionic species in that cell. For a fixed radiation field the ionization
parameter decreases as the density of the cell increases. A decrease in the
ionization parameter, implies a decrease in the effectiveness of the incident
radiation field.  This effect is seen Fig. \ref{fig:tpe}, where the relative
difference in cooling times of the gas cells under the presence and absence of
hot halo radiation fields is much smaller for the high density gas cells. For
example, the gas cell with $T_{start}=3 \times 10^7$K (solid curves) has a
three fold increase in the cooling time due to the hot halo radiation field
when it has a density of $\rm{n_H  = 0.001 \ cm^{-3}}$, but the cooling times
are practically identical when we increase the density of the cell to $\rm{n_H
= 0.1 \ cm^{-3}}$. This means that galaxies with lower mean densities will be
more amenable to  reduction in cooling rates due to local radiation fields.

\section{Results}
\label{sec:sim}

In order to study the effect of the hot halo radiation fields in a full
dynamical setting, we perform isolated galaxy simulations. Specifically, we
simulate the idealised clusters h14 and h15 (see Section \ref{sec:hhe} for more
details about the construction and properties of these clusters).

\subsection{Host Halo Quenching}
\label{sec:hhq}
We simulate two clusters of halo mass $10^{14} \ \rm{M}_\odot$ (h14) and
$10^{15} \ \rm{M}_\odot$ (h15).  The only physical processes considered are
gravity, hydrodynamics and gas cooling. A very simple prescription for self
shielding is also included by imposing the condition that all gas particles
with $\rm{n_{H}} \ge 0.1 \  \rm{cm^{-3}}$ \citep{Ceverino2010} have zero local
radiation flux. All the gas in the halo is assumed to have a metallicity of
$\rm{0.3Z_\odot}$, a value chosen to match the observational constraints from
nearby clusters \citep{Sato2008}.

\begin{figure}
\begin{center}
\includegraphics[scale=0.47]{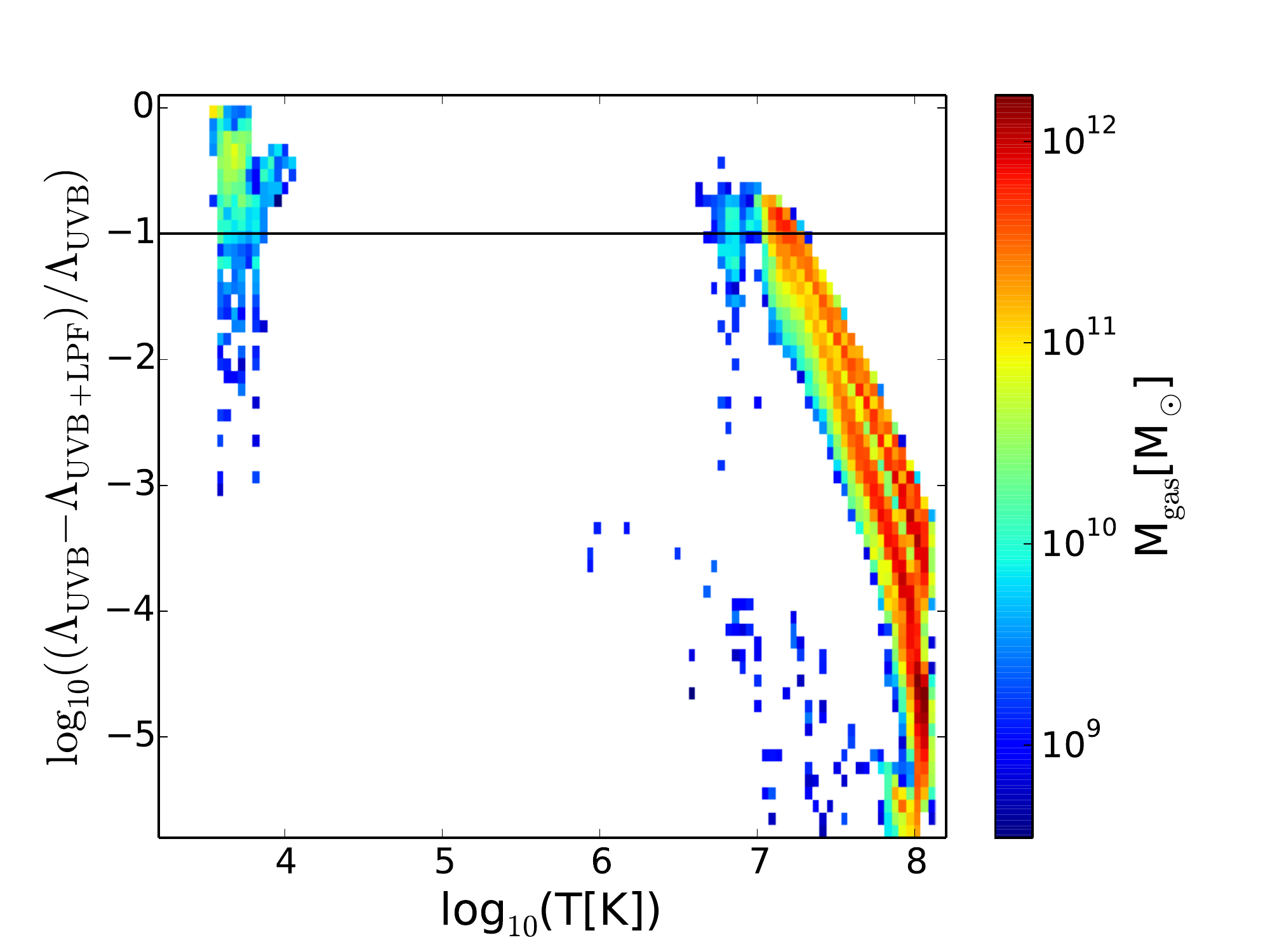}
\caption{A 2-D histogram of the amount of gas mass in the h15 cluster plotted
as a function of the change in cooling rate between the UVB and UVB+LPF
simulation runs and temperature at the end of $0.5$ Gyr of evolution.}
\label{fig:ctmain}
\end{center}
\end{figure}

Each cluster is simulated twice, once with gas cooling rates calculated in the
presence of the UV background radiation field at $z=0$ only (UVB) and once with
the hot halo radiation field turned on in addition to the UV background
(UVB+LPF). The simulations are run for $0.5$ Gyr.  The percentage change in the
amount of cold gas ($\rm{T \le 10^4K}$) formed between UVB and UVB+LPF
simulation for the h14 and h15 clusters is rather minimal ($\la 3 \%$). This is
because the virial temperature of clusters are high enough to fully thermally
ionize the gas, which reduces the effectiveness of the incident radiation
fields. This is seen more clearly in Fig. \ref{fig:ctmain}, which shows the
amount of gas mass as a function of temperature and the change in cooling rate
between the UVB and UVB+LPF simulations for the h15 cluster.  Most of the gas
in the halo lies in the region with $T> 10^{7.5}$K, where the change in cooling
rate is pretty minimal.  There is a $\ga 10\%$ (denoted by the solid horizontal
black line in the Fig. \ref{fig:ctmain}) decrease in the cooling rate of the
UVB+LPF simulation only in the region where $T<10^{7.5}$K. However, the amount
of gas in this region of the plot is only a small fraction of the total gas
mass. There is in fact a wide region between $10^4$K and $10^6$K, devoid of
gas, because the cooling time in this region is much lower than the dynamical
time of the system.  Below $10^4$K the radiation field does have a large
effect, but gas at this temperature will enter the modified equation of state
according to the ISM model used in our simulations \citep{Springel2003},
meaning that it has already been earmarked for star formation.  Therefore,
changing the cooling of this gas will not reduce the star formation rate in our
simulations.

\begin{figure}
\begin{center}
\includegraphics[scale=0.47]{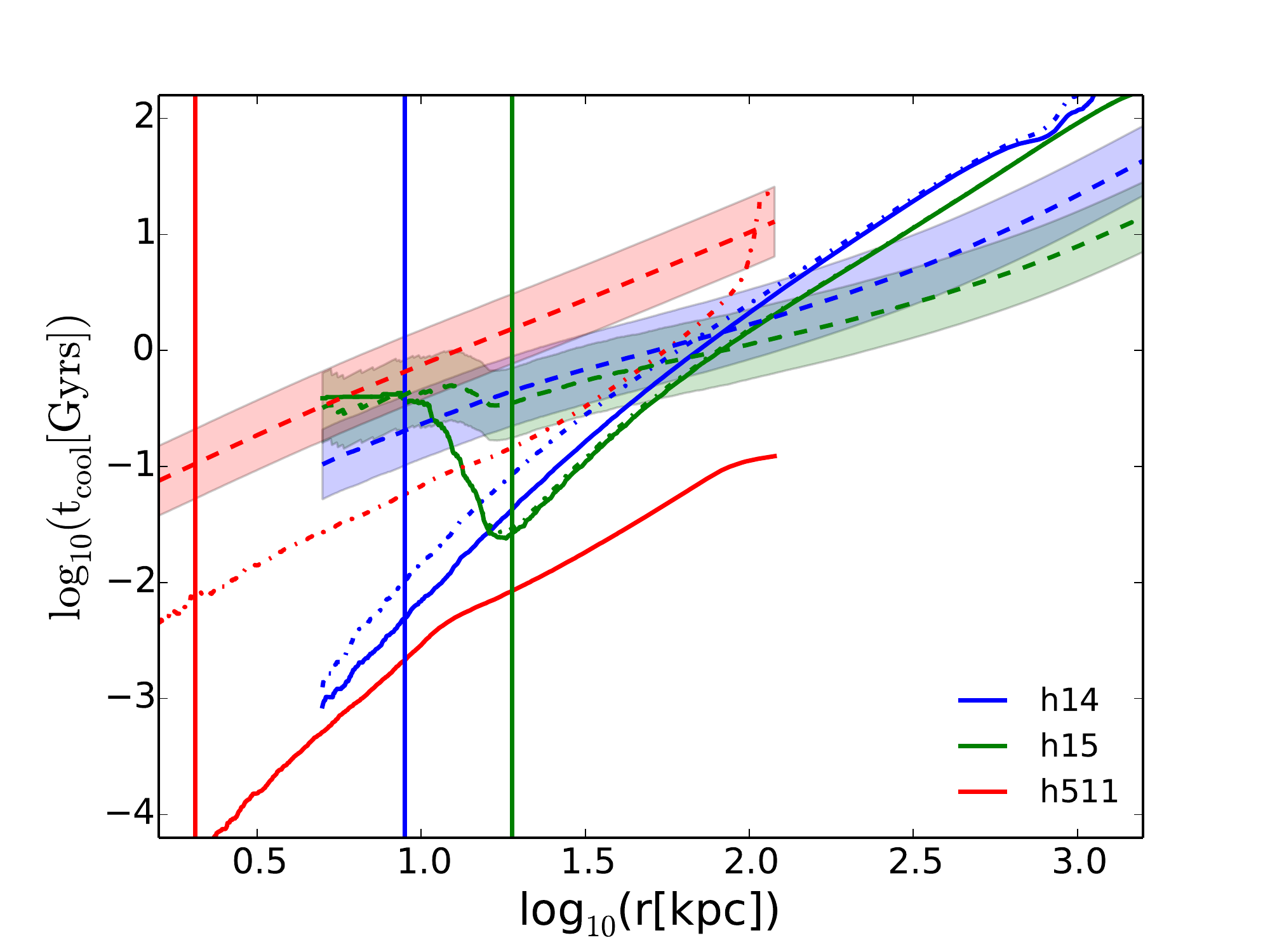}
\caption{Cooling time plotted as a function of radius in the h14 (blue curves),
h15 (green curves) and h511 (red curves) simulations. The dashed lines are
$10\rm{t_{ff}}$, and the corresponding shaded regions are $5\rm{t_{ff}}$ to
$20\rm{t_{ff}}$. The solid lines represent the cooling rates in the UVB
simulation while the dot-dashed lines are the cooling rates in the UVB+LPF
simulations. The vertical lines are $r=2.5\epsilon$ above which the properties
of the galaxies are well resolved.}
\label{fig:cprof}
\end{center}
\end{figure}

Another way to quantify the effect of local radiation fields is to look at the
change in cooling timescales as a function of the free-fall time of the system.
\citet{Voit2014N} have shown that clusters with on-going star formation and/or
AGN activity have cooling times about the order of precipitation time scales in
clusters, which are empirically seen to occur when $\rm{t_{ff}< t_{cool} <
20t_{ff}}$ \citep{Gaspari2011, Gaspari2012}, whereas, clusters with no star
formation or AGN activity were shown to lie above this precipitation zone,
meaning some  mechanism, other than SF or AGN, is responsible for stabilising
the cooling in these haloes.  Fig.~\ref{fig:cprof} shows the gas cooling times
as a function of radii for both the h14 (blue curves) and h15 (green curves)
cluster simulations. The solid lines show the cooling times for the UVB
simulation and the dot-dashed lines show the cooling times for the UVB+LPF
simulation. These values are shown for the cluster after $0.5$ Gyr of evolution.
In the h15 cluster the UVB and UVB+LPF cooling times lie on top of one another,
however, the h14 cluster does show a modest increase, a factor of $2$, in the
cooling times in the center. There are a couple of reasons why the radiation
field is marginally more effective in the h14 cluster. One, the virial
temperature of the h14 cluster is lower, meaning that the photons are more
likely to interact with the gas. Two. the free fall times in the h14 cluster
are slightly longer, meaning a lower central density, which in turn decreases
the density of the gas. This decreased gas density increases the cooling times
and will also be more amenable to reduction in cooling rates due to local
radiation fields (see Section  \ref{sec:tpe}).  However, in neither cluster is
the effect of the radiation field strong enough to increase the cooling
timescales above the precipitation timescales of the system. 

We find that the low density gas present in the outskirts of the cluster has
too high a temperature for the radiation field to effectively couple with it. On
the other hand, the relatively cold gas, with large photon interaction cross
sections, are at too high a density, thereby reducing the effect of the
radiation fields. We therefore conclude that hot halo radiation fields cannot
significantly impact star formation in massive clusters.

\subsection{Satellite Quenching}
\label{sec:enqn}
We have shown in Section \ref{sec:hhe} and Section \ref{sec:tpe}, that the
cooling rate of warm-hot gas ($10^4 - 10^6$K) in clusters is significantly
reduced in the presence of hot halo radiation fields.  It follows that in
systems where the dynamical timescales are similar to the cooling time of the
warm-hot gas, the effect of the radiation fields will be much more pronounced.
Low mass galaxies ($\rm{M_{halo} \la 10^{12} M_\odot}$) have  virial
temperatures in the range of $10^4-10^6$K and the dynamical timescales of the
system are of the order of $\sim 100$ Myrs. This means that a hot halo
radiation field, such as the one due to the ICM of the h15  cluster, will
change the dynamics of cooling flows in low mass systems. Therefore, satellite
galaxies within the cluster will be adversely affected by this ``radiation quenching`` mechanism. 

\begin{figure}
\begin{center}
\includegraphics[scale=0.47]{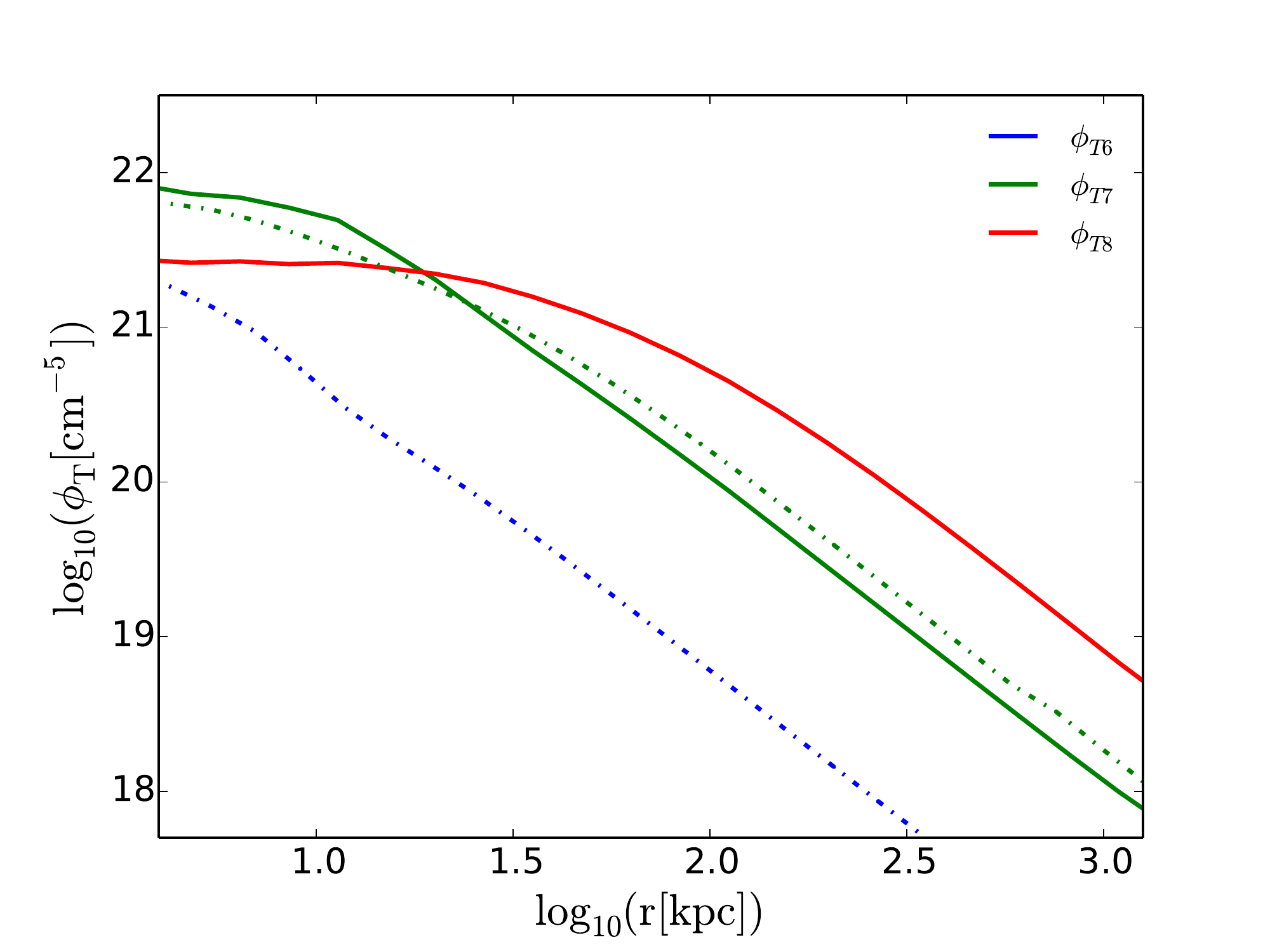}
\caption{Radiation field intensity plotted as a function of radius for the h14
(dot-dashed curves) and h15 (solid curves) clusters separated into the three
temperature components $\rm{\phi_{T6}}$  (blue curves),   $\rm{\phi_{T7}}$
(green curves)   $\&$   $\rm{\phi_{T8}}$  (red  curves).}
\label{fig:hhcrad}
\end{center}
\end{figure}

To test this scenario, we construct a galaxy of mass $5 \times 10^{11} \
\rm{M}_\odot$, labelled h511, in the same way as described in Section
\ref{sec:hhe}. The properties of this galaxy are given in Table
\ref{tab:clprop}. The virial temperature of this halo is $10^{5.8}$K. As shown
in Section \ref{sec:hhe},  the radiation field of the cluster varies with the
distance from the center (Fig. \ref{fig:spectra_hh}). Therefore the position of
the satellite within the cluster will determine how effective the host
radiation fields can be.  Fig. \ref{fig:hhcrad}, shows  flux in each component
of the hot halo radiation field, ($\rm{\phi_{T6} \ blue \ curve, \ \phi_{T7} \
green \ curve, \ \phi_{T8} \ red \ curve}$), as a function of distance from the
center of the corresponding cluster  (h14 - dot dashed curves  and h15 - solid
curves). 
The virial temperature of the h15 cluster is $\sim 10^8$K, therefore the amount
of gas in the $10^{5.5}<T\le10^{6.5}$K range is negligible and hence the
$\rm{\phi_{T6}}$ component is sub-dominant. Whereas, the virial temperature of
the h14 cluster is $\sim 10^7$K, hence the $\rm{\phi_{T8}}$ component is
non-existent. 

We place the h511 halo at a distance of $100$ kpc (blue curves) and $300$ kpc
(green curves) in both the h14 (dot dashed curves) and h15 (solid curves)
clusters and simulate the galaxy for $0.5$ Gyrs with the host halo radiation
field turned on. 

\begin{figure}
\begin{center}
\includegraphics[scale=0.47]{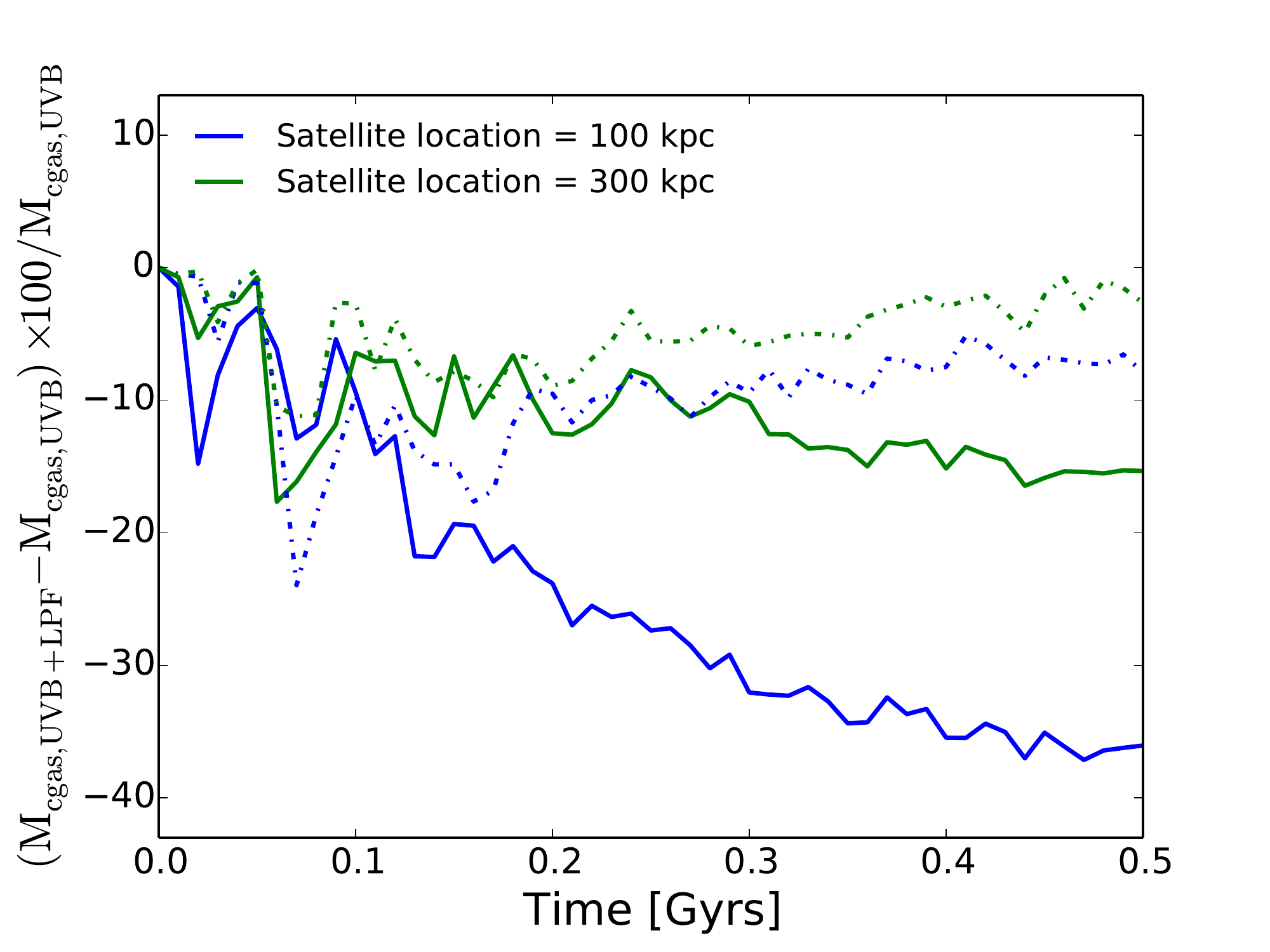}
\caption{Percentage change in the amount of cold gas ($T<10^4$K) formed between
the UVB+LPF and UVB simulations as a function of time in the h511 galaxy
simulation when placed at $100$ kpc (blue curves) and $300$ kpc (green curves)
from the center of the h15 (solid curves) and h14 (dot dashed curves)
clusters.} 
\label{fig:ccool}
\end{center}
\end{figure}

There is $\sim 40\%$ reduction in the amount of cool ($\rm{T<10^4K}$) gas
formed (Fig. \ref{fig:ccool}), if the satellite is placed at a distance of
$100$ kpc from the center of the h15 cluster compared to the field and the
effect reduces to $\sim 20 \%$ at the distance of $300$ kpc from the center.
The effect is smaller if the galaxy is placed in the h14 halo with the
reduction in amount of cool gas about $ 10 \%$. 

\begin{figure}
\begin{center}
\includegraphics[scale=0.47]{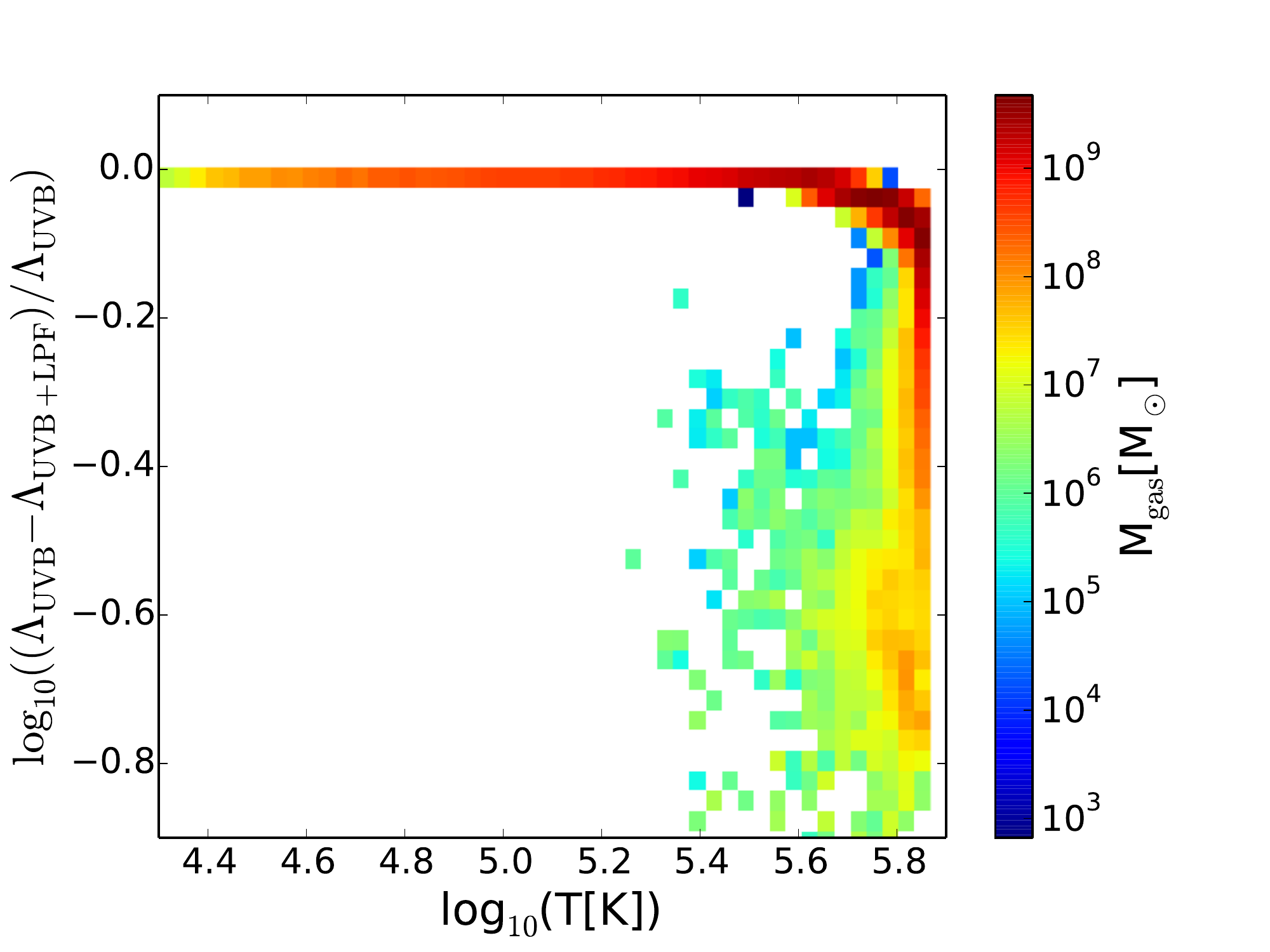}
\caption{A 2-D histogram of the amount of gas mass in the h511 halo plotted as
a function of the change in cooling rate between the UVB and UVB+LPF simulation
runs and temperature at the end of $0.5$ Gyr of evolution. The h511 galaxy is
assumed to be at a distance of $100$ kpc from the center of the h15 cluster.}
\label{fig:hist}
\end{center}
\end{figure}

Now the question arises as to why we see an affect on low mass galaxies. To
answer this we plot the gas mass of the satellite as a function of the
fractional change in the cooling rate of gas between the UVB and UVB+LPF
simulations and its temperature, when placed at a distance of $100$ kpc from
the center of the h15 cluster  (Fig.  \ref{fig:hist}). The cooling rates of
almost all the gas in the satellite is severely reduced in the presence of the
host radiation field. This in turn causes a reduction in the amount of cooling
flow and hence a reduction in the amount of cold gas formed. In contrast the
amount of gas with reduced cooling rates due to the hot halo radiation field
was minimal for the host cluster (see Fig. \ref{fig:ctmain}). 

Fig. \ref{fig:cprof} shows the cooling time of the gas in h511 halo, at a
distance of $100$ kpc from the center of the h15 cluster, as a function of
radii (red curves). In the absence of local radiation fields the cooling time
scales (red, solid curve) are $\sim 10^{-4} \ \rm{t_{prep}}$ (red, dashed curve
and shaded region), implying a high star formation rate.  The radiation fields
increase the cooling times by a factor of about $100-1000$ (red, dot-dashed
curve), making them only a order of magnitude lower than the precipitation
timescales. This huge effect is due to the fact that, unlike clusters, most of
the gas in the satellite is at exactly the temperatures at which the ionization
cross section in maximum ($< 10^6$K), and at low densities, as evidenced by the
constantly higher free fall times in the satellite. These increased cooling
times will lead to a drastic reduction in star formation rate of the galaxy,
just because of lower amounts of cooling flows (see RK14b). 

\subsubsection{Relative effectiveness of the radiation quenching mechanism}

Radiation quenching complements other environmental quenching mechanisms such
as ram pressure \citep{Gunn1972, MCC2008, Font2008, Kang2008, Weinmann2010}
and  tidal stripping  \citep{Taylor2001}. Therefore, it is important to
understand the impact of radiation quenching in the presence of these other environmental quenching mechanisms.  \citet{MCC2008} have found that for satellite galaxies
with typical structural and orbital parameters, up to $30 \%$ of the initial
hot halo gas can remain with the galaxy for up to $10$ Gyr. During this phase
it was assumed that the halo gas will continually cool and replenish the disk
and sustain star formation in the satellite. However, our results show that the
host halo radiation fields will reduce the star formation in these satellites
to a certain extent, even before they are completely stripped of gas. 

To test this scenario we run
a non-cosmological merger simulation of a $2 \times 10^{12} \ \rm{M_\odot}$ halo falling
into a $10^{15} \  \rm{M_\odot}$ cluster.  We construct a high resolution
version of the h15 cluster, where the DM distribution is sampled by $2.5$
million particles and the gas distribution sampled by $5$ million particles
(h15hr). The galaxy of mass $2 \times 10^{12} \ \rm{M_\odot}$ (h212lr) is
sampled with $5000$ DM particles and  $10000$ gas cells. These galaxies are
constructed in the same way as described in Section \ref{sec:hhe} and the
properties of them are listed in Table \ref{tab:clprop}.  The h212lr satellite
galaxy is placed at the virial radius of the h15hr cluster with the velocity of
the satellite given by the relations outlined in  \citet{Benson2005}. The system
is then evolved for $1$ Gyr with (UVB+LPF) and without (UVB) the hot halo
radiation fields turned on. In addition to gravity, hydrodynamics and gas
cooling, we also turn on star formation and stellar feedback according to the
models described in \citet{Vogelsberger2013}. 

\begin{figure}
\begin{center}
\includegraphics[scale=0.42]{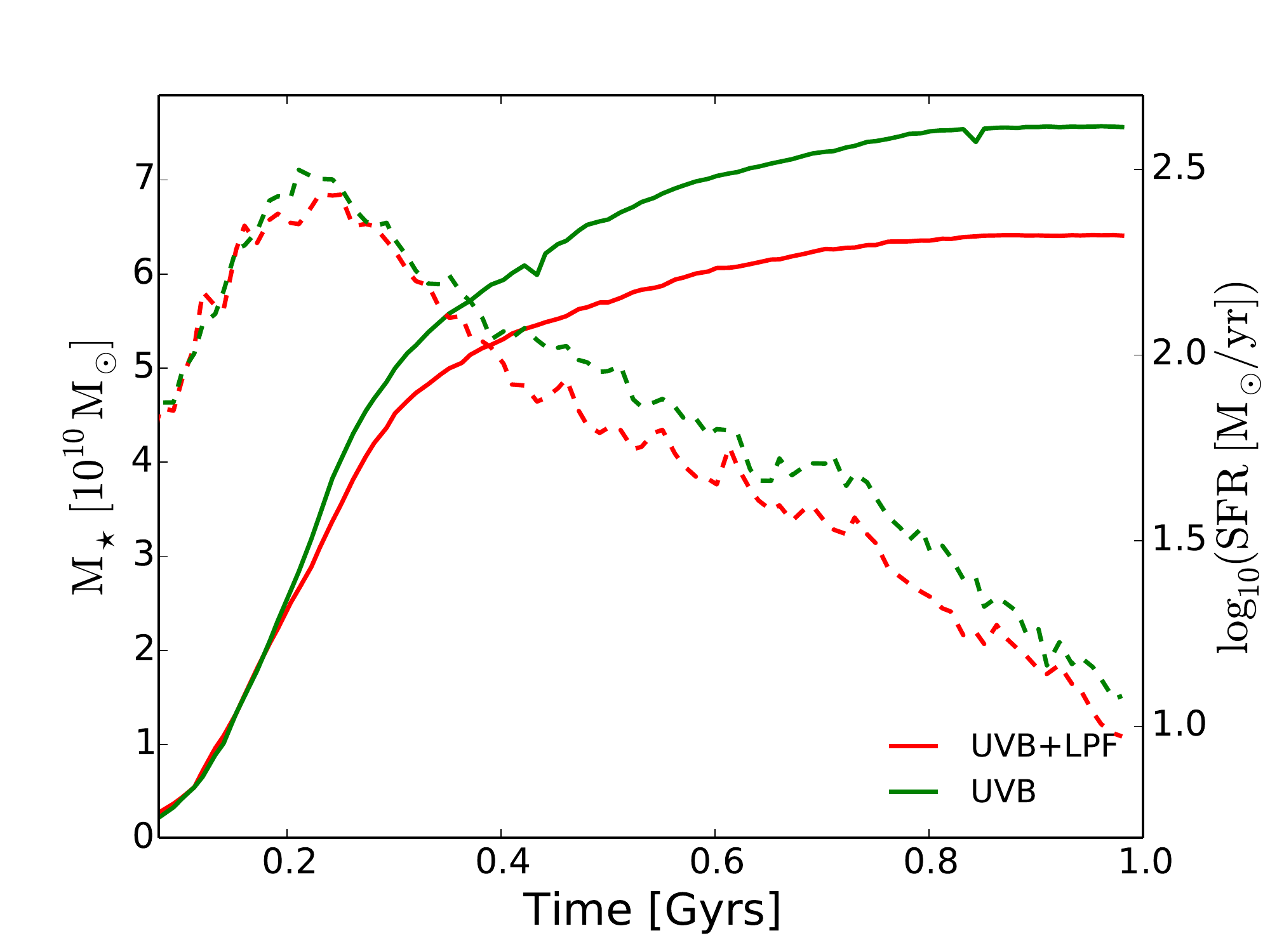}
\caption{The amount of stellar mass formed (solid curves) and the instantaneous
star formation rate (dashed curves) in the UVB+LPF (red curves) and UVB (green
curves) runs as a function of simulation time in the h212lr satellite halo,
when orbiting the h15hr cluster. }
\label{fig:stm}
\end{center}
\end{figure} 

Fig. \ref{fig:stm} shows the amount of stellar mass (solid curves) formed and the instantaneous star formation rates (dashed curves) in the
h212lr galaxy as it falls in and orbits around the h15hr cluster. The amount of
stellar mass formed after $1$ Gyr is about $20\%$ lower in the UVB+LPF run (red
curve) compared to the UVB run (green curve).  Additionally, the star formation
rate in the UVB+LPF simulation is also about $20-30\%$ lower. These results
suggest that even in the presence of other environmental quenching mechanisms,
radiation quenching plays a significant role in reducing star formation and
stellar mass in cluster environments. Therefore, they can in principle account
for the reduced star formation rates of  galaxies in cluster environments
\citep{Poggianti1999, Mel2008} and also explain why local quiescent galaxies
are observed to be quenched by a ''strangulation`` mechanism rather than by sudden
outflows of gas due to SNe feedback or ram pressure stripping \citep{Peng2015N,
Lu2015}.

\section{Conclusions}
\label{sec:conc}
In this paper, we present first simulations of galaxy formation which include
the impact X-ray emission from binaries and hot ICM on gas cooling rates using
the {\sc AREPO} code.  

We find the observed X-ray flux from XRBs to be sub-dominant compared to other
stellar radiation sources, minimally affecting the gas cooling rates in
galaxies.  However, hot gas emission from clusters has a high photon flux all
the way to the virial radius of the halo. This is because the ICM is an
extended source, therefore the bremsstrahlung emission from the ICM will not
fall off as rapidly as radiation from centralised sources like stars. The gas
cooling  rates are substantially reduced, especially in the
$10^4\rm{K}<T<10^6$K regime, under the influence of these radiation fields.

We test our model in a full dynamical setting by simulating idealised clusters
of mass $10^{14} \ \rm{M_\odot}$ and $10^{15} \ \rm{M_\odot}$.  The difference
between the amount of cold gas  ($\rm{T}< 10^4$K) formed between the
simulations with and without local radiation feedback is minimal, because the
cooling flows are not affected by the local radiation sources in clusters. This
is because the low density gas present in the outskirts of the cluster is at 
too high a temperature for the radiation field to effectively couple with it. On
the other hand, the relatively cold gas, with large photon interaction cross
sections, are at too high a density, which reduces the effectiveness of the
radiation fields. This means that, even though there is enough energy in the
form of radiation to quench gas cooling in clusters, the  coupling between the
gas and the radiation field is too weak. We conclude that the radiation fields
considered in this paper will not have a major effect on star formation rates
in massive galaxies.  

The cooling rate of relatively cold gas ($T<10^6$K) on the other hand is
reduced considerably by the hot-halo radiation field. The change in the cooling
rate can be as large as the $30$ times the fiducial value. This can in
principal change the dynamics of cooling flows in the systems where the
dynamical timescales are comparable to the cooling timescales of gas. This
implies that relatively low mass satellites ($\rm{M_{halo}  \la 10^{12}
M_\odot}$) will experience quenching due to the radiation field of the host
halo in cluster environments.  We tested this idea by simulating a low mass
galaxy ($\rm{M_{halo} = 5 \times 10^{11} \ M_\odot}$) in the presence of host
halo radiation fields. We placed the satellite galaxy at a various distances
from the center of the cluster and assumed that all gas in the satellite sees a
host halo radiation field corresponding to that distance.  These simulations
show an apparent reduction in the amount of cold gas formed when the host halo
radiation field is turned on. At a distance of $100$ kpc from the center of the
halo, the satellite has a $\sim 40\%$ reduction in cold gas mass in $0.5$ Gyrs.
This reduces to about $20\%$ at $300$ kpc and reduces further at larger radii.
This effect is due to the fact that, unlike clusters, most of the gas in the
satellite is at exactly the temperatures at which the ionization cross section
is largest ($T< 10^6$K), and have low enough densities, to be highly influenced
by the radiation field.

We also compared the effectiveness of radiation quenching to other
environmental quenching mechanisms (e.g., ram pressure stripping) by running a galaxy merger simulation of a $2 \times 10^{12} \ \rm{M_\odot}$ halo
falling into $10^{15} \ \rm{M_\odot}$ halo. The UVB+LPF simulation shows a
$20\%$ reduction in the stellar mass of the satellite after $1$ Gyr with
respect to the UVB-only simulation. We also find a reduction of $\sim 20-30 \%$
in star formation rates at late times. These results indicate that even in the
presence of other environmental quenching mechanisms, the hot halo radiation
fields still play a significant role in quenching galaxies in cluster
environments.
 
We note that these results are preliminary and we plan to extend this initial
study to a full cosmological simulation and quantify the effectiveness of the
local radiation fields on satellite quenching in a wide range of satellite and
host galaxy masses.

\section*{Acknowledgements}
The simulations were performed on the joined MIT-Harvard computing cluster
supported by MKI and FAS and the Milky Way supercomputer, funded by the
Deutsche Forschungsgemeinschaft (DFG) through Collaborative Research Center
(SFB 881) `The Milky Way System’ (subproject Z2), hosted and cofunded by the
J\"ulich Supercomputing Center (JSC). We greatly appreciate the contributions
of these computing allocations.  GSS, VS and AVM acknowledge support from SFB
881 'The Milky Way System’ (subproject A1) of the German Research Foundation
(DFG). JFH acknowledges generous support from the Alexander von Humboldt
foundation in the context of the Sofja Kovalevskaja Award. The Humboldt
foundation is funded by the German Federal Ministry for Education and Research.
\bibliographystyle{mn2e}
\bibliography{references}

\clearpage

\end{document}